%% file: paper.tex
\RequirePackage{fix-cm}
\documentclass{article}          
\pdfoutput=1

\usepackage{bm} 
\usepackage{lmodern}
\usepackage{moreverb}

\usepackage[colorlinks,bookmarksopen,bookmarksnumbered,citecolor=blue,urlcolor=blue]{hyperref}

\usepackage{makeidx}         
\usepackage{graphicx}        
\usepackage{multicol}        
\usepackage{url}
\usepackage{tabularx,booktabs}
\usepackage{setspace}
\renewcommand{\PackageWarningNoLine}[2]{}
\newcommand{\tensor}[1]{\mathbf{#1}}
\newcommand{\symbtensor}[1]{\boldsymbol{#1}}
\renewcommand{\vec}[1]{\mathbf{#1}}
\newcommand{\symbvec}[1]{\boldsymbol{#1}}
\newcommand{\Rvec}[1]{\underline{#1}} 
\newcommand{\dd}{\,\mathrm{d}}

\usepackage{amsmath,amssymb}
\usepackage{amsfonts}
\usepackage{graphicx}
\usepackage{color}
\usepackage{wasysym}
\usepackage{relsize}

\usepackage{geometry}
\usepackage{fancyhdr}
\def\keywords{\vspace{.5em}
{\textit{Keywords}:\,\relax%
}}

\usepackage{authblk}
\title{\bf Classical and all-floating FETI methods for the simulation of
arterial tissues}

\author{Christoph~M.~Augustin$^{1,3,}$\thanks{\ \texttt{christoph.augustin@medunigraz.at};
Corresponding author}\ ,
Gerhard~A.~Holzapfel$^{2}$, Olaf~Steinbach$^1$
}
\affil{\small$^1$Institute of Computational Mathematics,
Graz University of Technology,
\mbox{Steyrergasse 30, 8010 Graz, Austria}\break
$^2$Institute of Biomechanics, Center of Biomedical Engineering,
Graz University of Technology, \mbox{Kronesgasse 5}, 8010 Graz, Austria \break
$^3$Institute of Biophysics, Medical University of Graz,
Harrachgasse 21, 8010 Graz, Austria \break
\vspace*{-5mm}}
\date{}

\begin{document}

\maketitle
{\small This is the peer reviewed version of the following article:
Augustin, C. M., Holzapfel, G. A. and Steinbach, O. (2014),
\emph{Classical and all-floating FETI methods for the simulation of arterial tissues.}
Int. J. Numer. Meth. Engng., 99: 290--312. doi: 10.1002/nme.4674,
which has been published in final form at
\url{onlinelibrary.wiley.com/doi/10.1002/nme.4674/abstract}.
This article may be used for non-commercial purposes in
accordance with \href{http://olabout.wiley.com/WileyCDA/Section/id-820227.html#terms}
{Wiley Terms and Conditions for Self-Archiving}.}\\[2ex]
\begin{abstract}
High-resolution and anatomically realistic computer models of
biological soft tissues play a significant role in the
understanding of the function of cardiovascular components in
health and disease. However, the computational effort to handle
fine grids to resolve the geometries as well as sophisticated
tissue models is very challenging. One possibility to derive a
strongly scalable parallel solution algorithm is to consider
finite element tearing and interconnecting (FETI) methods. In this
study we propose and investigate the application of FETI methods
to simulate the elastic behavior of biological soft tissues. As
one particular example we choose the artery which is -- as most
other biological tissues -- characterized by anisotropic and
nonlinear material properties. We compare two specific approaches
of FETI methods, classical and all-floating, and investigate the
numerical behavior of different preconditioning techniques. In
comparison to classical FETI, the all-floating approach has not
only advantages concerning the implementation but in many cases
also concerning the convergence of the global iterative solution
method. This behavior is illustrated with numerical examples. We
present results of linear elastic simulations to show convergence
rates, as expected from the theory, and results from the more
sophisticated nonlinear case where we apply a well-known
anisotropic model to the realistic geometry of an artery. Although
the FETI methods have a great applicability on artery simulations
we will also discuss some limitations concerning the dependence on
material
parameters.
\end{abstract}
\keywords{artery, biological soft tissues, all-floating FETI,
parallel computing}
\section{Introduction}
The modeling of hyperelastic materials is realized
by using a strain--energy function $\Psi$. For a comprehensive
overview and the mathematical theory on elastic deformations, see,
e.g.,
\cite{Augustin_ciarlet1988,Augustin_holzapfel2000a,Augustin_marsden1994,Augustin_ogden1997}.
A well established model for arterial tissues was introduced by
Holzapfel et al.~\cite{Augustin_holzapfel2000b,
Augustin_holzapfel2010a}. This model was further developed and
enlarged to collagen fiber dispersion in
\cite{Augustin_holzapfel2010a,Augustin_gasser2006,Augustin_Holzapfel2008}; see \cite{Augustin_holzapfel2010b} for
the modeling of residual stresses in arteries which play also an
important role in tissue engineering. An adequate model for the
myocardium can be found in \cite{Augustin_holzapfel2009}. The fine
mesh structure to model cardiovascular organs normally results in
a very large number of degrees of freedom. The combination with
the high complexity of the underlying partial differential
equations demands fast solution algorithms and, conforming to
up--to--date computer hardware architectures, parallel methods.
One possibility to achieve these specifications is to use domain
decomposition (DD) methods which acquired a lot of attention in
the last years and resulted in the development of several
overlapping as well as non--overlapping DD methods, see
\cite{Augustin_DDM.org}. They all work according to the same
principle: the computational domain $\Omega_0$ is
subdivided into a set of (overlapping or non--overlapping)
subdomains $\Omega_{0,i}$. DD algorithms now
decompose the large global problem into a set of smaller local
problems on the subdomains, with suitable transmission or
interface conditions. This yields a natural parallelization of the
underlying problem. In addition to well established standard DD
methods, other examples for more advanced domain decomposition
methods are hybrid methods \cite{Augustin_steinbach2003}, mortar
methods
\cite{Augustin_bernardi1994,Augustin_maday1989,Augustin_wohlmuth2000}
and tearing and interconnecting methods
\cite{Augustin_farhat1991}.

In this paper we focus on the finite element tearing and interconnecting
(FETI) method where the strategy is to decompose
the computational domain into a finite number of non--overlapping subdomains.
Therein the corresponding local problems can be handled efficiently by direct
solvers. The reduced global system, that is related to discrete Lagrange
multipliers on the interface, is then solved with a parallel
Krylov space method to deduce the desired dual solution. This is, in the case
of elasticity, the boundary stress and
subsequentely, in a postprocessing step, we
compute the primal unknown, i.e. the displacements, locally. For the global
Krylov space method, such as the conjugate gradient (CG) or the
generalized minimal residual (GMRES) method, we need to have a suitable
preconditioning technique. Here we consider a simple lumped
preconditioner and an almost  optimal Dirichlet preconditioner,
as proposed by Farhat et al.~\cite{Augustin_farhat1994a}.

A variant of the classical FETI method is the all-floating tearing and
interconnecting approach (AF-FETI) where, in contrast to the classical
approach, the Dirichlet boundary acts as a part of the interface. It was
introduced independently for the boundary element method by Steinbach and Of
\cite{Augustin_of2006,Augustin_of2009} and as the Total-FETI (TFETI) method
for finite elements by Dost\'al et al.~\cite{Augustin_dostal2006}.
This approach shows advantages in the implementation and, due to mapping
properties of the involved operators, improves the convergence of the
global iterative method for the considered problems. This behavior is
illustrated with numerical examples, which are -- to the best of our
knowledge -- the first application of all-floating FETI method to nonlinear
and anisotropic biological materials.

An essential part of FETI methods is solving the local
subproblems. Challenges occur with so-called \textit{floating
subdomains} which have no contribution to the Dirichlet boundary.
These cases correspond to local Neumann problems and the solutions
are -- in the case of elasticity -- only unique up to the six
rigid body modes. For classical FETI it can happen
that the kernel of the local operator is non-trivial and its
dimension is lower than six. The problem to identify these kernels
reliably causes trouble. One possibility to overcome this trouble
is a modification of the classical approach, the dual-primal FETI
(FETI-DP) method, cf.~Farhat et al.~\cite{Augustin_farhat2001}
and Klawonn and Widlund \cite{Augustin_klawonn2001a}. In this
variant some specific primal degrees of freedom are fixed. This
yields solvable systems for all subdomains. Choosing the primal
degrees of freedom may be very sophisticated
\cite{Augustin_klawonn2005}. This approach was already applied to
model arterial tissues using FETI-DP by Klawonn and
Rheinbach \cite{Augustin_klawonn2010,Augustin_rheinbach2009},
Brands et al. \cite{Augustin_brands2008}, Balzani et al.
\cite{Augustin_balzani2010, Augustin_balzani2012} and Brinkhues et
al. \cite{Augustin_brinkhues2013}.
Note that for all-floating FETI the identification of the kernel
of the local operators is no problem at all, since we treat all
subdomains as floating subdomains, and hence have a kernel equal
to six for all local operators. Moreover the resulting local
systems are typically better conditioned than those arising in the
FETI-DP approach, see Brzobohat{\`y} et al.
\cite{Augustin_brzobohaty2011}. All-floating FETI was used to
model myocardial tissue in the preliminary work
\cite{Augustin_augustindd20}.

Both the classical FETI method, as well as all-floating FETI,
need the construction of a generalized inverse matrix. This may be
achieved using direct solvers with a sparsity preserving
stabilization, see, e.g. \cite{Augustin_brzobohaty2011}, or
stabilized iterative methods. For a mathematical analysis of FETI
methods including convergence proofs for the classical one-level
FETI method, see, e.g.,
\cite{Augustin_klawonn2001a,Augustin_klawonn2000,Augustin_mandel1996}.
\section{Modeling Arterial Tissues}\label{Augustin_sec:model}
The deformation of a body $\mathcal{B}$ is described by a function
$\boldsymbol{\phi}:\Omega_0\to\Omega_t$ with the \textit{reference
configuration} $\Omega_0 \subset \mathbb{R}^3$ at time $t=0$ and
the \textit{current configuration} $\Omega_t$ at time $t>0$. With
this we introduce the displacement field $\vec{U}$ in the
reference configuration and the displacement field $\vec{u}$ in
the current configuration,
\begin{equation}
 \vec{x} = \symbtensor{\phi}(\vec{X})
=\vec{X}+\vec{U}(\vec{X}) \in \Omega_t, \quad \vec{X} =
\symbtensor{\phi}^{-1}(\vec{x}) = \vec{x}-\vec{u}(\vec{x}) \in
\Omega_0,
\end{equation}
and the deformation gradient as, see, e.g.,
\cite{Augustin_holzapfel2000a},
\begin{equation}
 \tensor{F} =\operatorname{Grad}\symbvec{\phi}(\vec{X}) =
 \tensor{I}+\operatorname{Grad}\vec{U}.
\end{equation}
Moreover, we denote by $J=\det \tensor{F}>0$ the Jacobian of
$\tensor{F}$ and by $\tensor{C}=\tensor{F}^\top \tensor{F}$ the
right Cauchy--Green tensor. For later use, to model the nearly
incompressible behavior of biological soft tissues, we introduce
the following split of the deformation gradient in a volumetric
and an isochoric part, compare Flory \cite{Augustin_flory1961},
i.e.
\begin{equation}\label{Augustin_eq:decompositionFlory}
      \tensor{F}=J^{1/3}\overline{\tensor{F}},
      \quad \text{with } \det\overline{\tensor{F}}=1.
\end{equation}
Consequently, this multiplicative split can be applied to other tensors
such as the right Cauchy--Green tensor. Thus
\begin{equation}
      \tensor{C}=J^{2/3}\overline{\tensor{C}}, \quad \text{with }
      \overline{\tensor{C}}=\overline{\tensor{F}}^\top\overline{\tensor{F}}\ \text{ and }
      \det\overline{\tensor{C}}=1.
\end{equation}
As a starting point for the modeling of biological soft tissues
the stationary equilibrium equations in the current configuration
are considered to find a displacement field $\vec{u}$ according to
\begin{equation}\label{Augustin_eq:current}
\operatorname{div} \symbtensor{\sigma} (\vec{u},\vec{x}) + \vec{b}_t(\vec{x}) =
\vec{0} \quad \mbox{for} \; \vec{x} \in \Omega_t ,
\end{equation}
where $\symbtensor{\sigma} (\vec{u},\vec{x}) $ is the Cauchy stress tensor and
$\vec{b}_t(\vec{x}) $ is the body force at time $t$.

In addition, we incorporate boundary conditions to describe displacements or
normal stresses on the boundary $\Gamma_t=\partial \Omega_t$, which is
decomposed into disjoint parts such that
$\partial\Omega_t=\overline{\Gamma}_{t,\mathrm{D}} \cup \overline{\Gamma}_{t,\mathrm{N}}$.
Dirichlet boundary conditions on $\Gamma_{t,\mathrm{D}}$ correspond to a given
displacement field $\vec{u}=\vec{u}_\mathrm{D}(\vec{x})$, while Neumann boundary
conditions on $\Gamma_{t,\mathrm{N}}$ are identified physically with a given surface
traction $\symbtensor{\sigma}(\vec{u},\vec{x})\,\vec{n}_t(\vec{x})
=\vec{g}_t(\vec{x})$, where $\vec{n}_t(\vec{x})$ denotes the exterior normal
vector at time $t$.

The equilibrium equations and the boundary conditions may also be formulated
in terms of the reference configuration, i.e.
\begin{eqnarray}\label{Augustin_eq:reference}
\operatorname{Div}\tensor{F}\tensor{S}(\vec{U},\vec{X})+\vec{b}_0(\vec{X})&=&\vec{0}
\hspace*{13.7mm} \text{for }  \vec{X}\in\Omega_0 ,\\
\vec{U}(\vec{X})&=&\vec{U}_\mathrm{D}(\vec{X}) \hspace*{4.5mm}
\text{for } \vec{X}\in\Gamma_{0,\mathrm{D}} \label{Augustin_eq:Dirichlet},  \\
\tensor{F}\tensor{S}(\vec{U},\vec{X}) \vec{N}_0(\vec{X})&=&
\vec{G}_0(\vec{X}) \hspace*{5mm}
\text{for } \vec{X}\in\Gamma_{0,\mathrm{N}}  , \label{Augustin_eq:Neumann}
\end{eqnarray}
where $\tensor{S}$ is the second Piola--Kirchhoff tensor and
$\vec{b}_0(\vec{X}) $ is the body force at time $t=0$. In order to formulate
the boundary conditions we introduce a prescribed displacement field
$\vec{U}_\mathrm{D}(\vec{X})$,  the exterior normal vector $\vec{N}_0(\vec{X})$ and
the surface traction $\vec{G}_0(\vec{X})$ in the reference configuration.

Considering the study of the properties of soft biological soft
tissues we have to deal with a nonlinear relationship between
stress and strain, with large deformations and an anisotropic
material. Since linear elasticity models are not adequate for
treating such a complex behavior, we take a look at the more
general concept of nonlinear elasticity.

The nonlinear stress-strain response is modeled via a
constitutive equation that links the stress to a derivative of a
strain-energy function $\Psi$, representing the elastic stored
energy per unit reference volume. Derived from the Clausius--Duhem
inequality, see
\cite{Augustin_coleman1963,Augustin_truesdell1960}, we formulate
the constitutive equations as
\begin{equation}
  \symbtensor{\sigma} = 2J^{-1} \tensor{F} \frac{\partial
    \Psi(\tensor{C})}{\partial \tensor{C}} \tensor{F}^\top \quad\text{and}
    \quad  \tensor{S} = 2 \frac{\partial \Psi(\tensor{C})}{\partial \tensor{C}}.
\end{equation}
We make use of the Rivlin--Ericksen representation theorem
\cite{Augustin_rivlin1955} and its extension to anisotropic
materials, cf. \cite{Augustin_raoult2009}, to find a
representation of the strain-energy function $\Psi$ in terms of
the principal invariants of $\tensor{C}$.

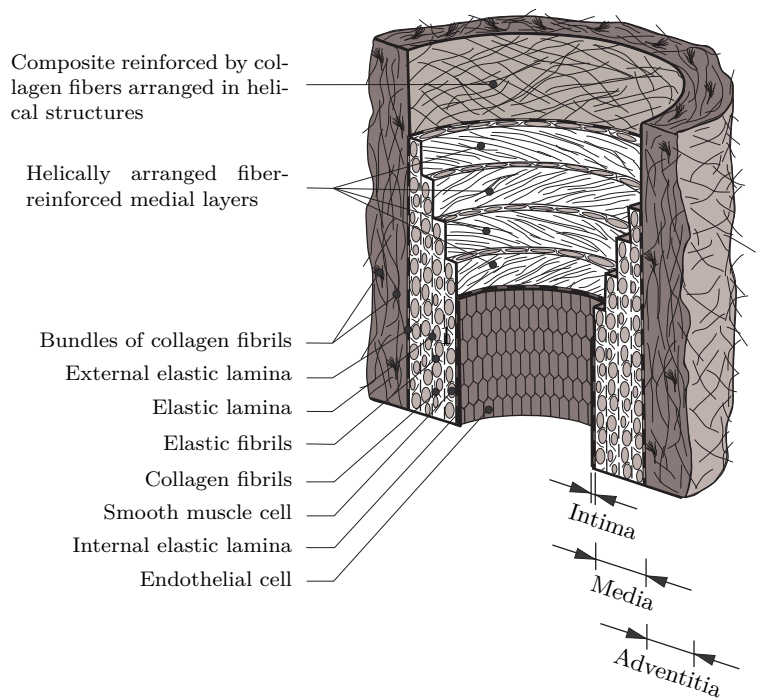
\begin{figure}[!b]
      \centering
      \def\svgwidth{60mm}\input{augustin_artery.tex}
       \caption{Diagrammatic model of the major components of a healthy
       elastic artery, from \cite{Augustin_holzapfel2000b}. The intima, the
       innermost layer is negligible for the modeling of healthy arteries, it
       plays a very important role in the modeling of diseased arteries, though.
       The two predominant directions of the collagen fibers in the media
       and the adventitia are indicated with black curves.}
       \label{Augustin_fig:arterialModel}
   \end{figure}

Arteries are vessels that transport blood from the heart to the
organs. In vivo the artery is a prestretched material under an
internal pressure load. Healthy arteries are highly deformable
composite structures and show a nonlinear stress-strain response
with a typical stiffening effect at higher pressures. Reasons for
this are the embedded collagen fibers which lead to an anisotropic
mechanical behavior of arterial walls. We denote by
$\vec{a}_{0,1}$ and $\vec{a}_{0,2}$ the predominant collagen fiber
directions in the reference configuration. An important
observation is that arteries do not change their volume within the
physiological range of deformation, hence they are treated as a
nearly incompressible material, see, e.g.,
\cite{Augustin_holzapfel2000b}. In this work we focus on the in
vitro passive behavior of the healthy artery, see
Fig.~\ref{Augustin_fig:arterialModel}.
To capture the nearly incompressibility condition we remember the
decomposition \eqref{Augustin_eq:decompositionFlory}, which yields an additive
split of the strain-energy function into a so-called volumetric and an
isochoric part, i.e.
\begin{equation}
    \label{Augustin_eq:decompositionStrain}
    \Psi(\tensor{C})=\Psi_{\text{vol}}(J)+\overline{\Psi}(\overline{\tensor{C}}).
\end{equation}
This procedure leads to constitutive equations in which the stress tensors are
also additively decomposed into a volumetric and an isochoric part, i.e.,
cf. \cite{Augustin_holzapfel2000a},
\begin{equation}\label{Augustin_eq:additiveStress}
  \symbtensor{\sigma} = p\tensor{I}+2J^{-1} \tensor{F} \frac{\partial
    \overline{\Psi}(\overline{\tensor{C}})}{\partial \tensor{C}} \tensor{F}^\top
    \quad\text{and}
    \quad  \tensor{S} = Jp\tensor{C}^{-1}+2 \frac{\partial
      \overline{\Psi}(\overline{\tensor{C}})}{\partial \tensor{C}}.
\end{equation}
Here, the scalar-valued hydrostatic pressure is defined as
\begin{equation}
  p:=\frac{\partial\Psi_\text{vol}(J)}{\partial J}.
  \label{Augustin_eq:hydrostaticPressure}
\end{equation}
To capture the specifics of this fiber-reinforced composite,
Holzapfel and Weizs{\"a}cker \cite{Holz1998a} and
Holzapfel et al.~\cite{Augustin_holzapfel2000b} proposed an
additional split of the strain-energy function into an isotropic
and an anisotropic part so that the complete energy
function $\Psi$ can be written as
\begin{equation} \label{Augustin_eq:isoaniso}
    \Psi(\tensor{C}) = \Psi_{\text{vol}}(J)+
    \overline{\Psi}_{\text{iso}}(\overline{\tensor{C}}) +
    \overline{\Psi}_{\text{aniso}}(\overline{\tensor{C}},\vec{a}_{1,0})+
    \overline{\Psi}_{\text{aniso}}(\overline{\tensor{C}},\vec{a}_{2,0}).
\end{equation}
Following the classical approach we describe the volume changing part by
\begin{equation}
  \Psi_{\text{vol}}(J)=\frac{\kappa}{2}(J-1)^2,
\end{equation}
where $\kappa>0$, comparable to the bulk modulus in linear
elasticity, serves as a penalty parameter to enforce the
incompressibility constraint.

To model the isotropic non-collagenous matrix
material the classical neo-Hookean model is used
\cite{Augustin_holzapfel2000a}. Thus
\begin{equation}\label{Augustin_eq:iso}
  \overline{\Psi}_\text{iso}(\overline{\tensor{C}}) =
    \frac{c}{2}(\overline{I}_1-3) ,
\end{equation}
where $c>0$ is a stress-like material parameter and
$\overline{I}_1=\operatorname{tr}(\overline{\tensor{C}})$ is the
first principal invariant of the isochoric part of the right
Cauchy--Green tensor. In \eqref{Augustin_eq:isoaniso},
$\overline{\Psi}_{\text{aniso}}$ is associated with the
deformation of the collagen fibers. According to
\cite{Augustin_holzapfel2000b}, this transversely isotropic
response is described by
\begin{align}
\overline{\Psi}_\text{aniso}(\overline{\tensor{C}},\vec{a}_{1,0})&=
\frac{k_1}{2k_2}\left\{\exp[k_2(\overline{I}_{4}-1)^2]-1\right\}
\label{Augustin_eq:aniso_1},\\
\overline{\Psi}_\text{aniso}(\overline{\tensor{C}},\vec{a}_{2,0})&=
\frac{k_1}{2k_2}\left\{\exp[k_2(\overline{I}_{6}-1)^2]-1\right\}
\label{Augustin_eq:aniso_2},
\end{align}
with the invariants
$\overline{I}_{4}:=\vec{a}_{1,0}\cdot(\overline{\tensor{C}}\vec{a}_{1,0})$,
$\overline{I}_{6}:=\vec{a}_{2,0}\cdot(\overline{\tensor{C}}\vec{a}_{2,0})$
and the material parameters $k_1$ and $k_2$, which are both
assumed to be positive. It is worth to mention that for the
anisotropic responses, \eqref{Augustin_eq:aniso_1} and
\eqref{Augustin_eq:aniso_2} only contribute for the cases
$\overline{I}_{4}>1$ or $\overline{I}_{6}>1$, respectively. This
condition is explained with the wavy structure of the collagen
fibers, which are regarded as not being able to support
compressive stresses. Thus, the fibers are assumed to be active in
tension ($\overline{I}_i > 1$) and inactive in
compression ($\overline{I}_i<1$). This assumption is not only
based on physical reasons but it is also essential for reasons of
stability, see Holzapfel et al. \cite{Augustin_holzapfel2004}.

The material parameters can be fitted to an experimentally
observed response of the biological soft tissue. Following
\cite{Augustin_holzapfel2000b} we use the material parameters
summarized in Table~\ref{Augustin_tab:param}.

\begin{table}
  \caption{Material parameters used in the numerical experiments; parameters
  taken from Holzapfel~et~al.~\cite{Augustin_holzapfel2000b}.}
  \centering
\small
\begin{tabular}{lll}
  \toprule
$c = 3.0$\,kPa$\hspace{0.8cm}$   & $k_1 =
2.3632$\,kPa$\hspace{0.8cm}$&
$k_2=0.8393$ (-)  \\
\bottomrule
\end{tabular}
\label{Augustin_tab:param}
\end{table}

Similar models can also be used for the description of other biological
materials, e.g., for the myocardium, cf. \cite{Augustin_holzapfel2009}.
\section{Finite Element Approximation}\label{Augustin_sec:FEM}
\subsection{Variational formulation of nonlinear elasticity problems}\label{Augustin_sec:varForm}
In this section we consider the variational formulation of the
equilibrium equations \eqref{Augustin_eq:current} and
\eqref{Augustin_eq:reference} with the corresponding Dirichlet and Neumann
boundary conditions. In particular, using spatial coordinates, the boundary value
problem \eqref{Augustin_eq:current} is formally equivalent to the variational equations
\begin{equation}
\langle \mathcal{A}_t(\vec{u}) , \vec{v} \rangle_{\Omega_t}
:=\int_{\Omega_t} \symbtensor{\sigma}(\vec{u}) :
\symbtensor{\varepsilon}(\vec{v}) \dd \vec{x}
= \int_{\Omega_t} \vec{b}_t \cdot \vec{v} \dd \vec{x} +
\int_{\Gamma_{t,\mathrm{N}}} \vec{g}_t \cdot \vec{v} \dd s_{\vec{x}} =:
\langle \vec{\mathcal{F}},\vec{v}\rangle_{\Omega_t},
\label{Augustin_eq:WeakCurrent}
\end{equation}
valid for a smooth enough tensor field $\symbtensor{\sigma}(\vec{u}):\overline{\Omega}_t\mapsto\mathbb{R}^{3\times 3}$ and all smooth enough vector fields
$\vec{v}:\overline{\Omega}_t\mapsto\mathbb{R}^3$, which vanish on $\Gamma_{t,\mathrm{D}}$, see, e.g., \cite[Theorem~2.4-1]{Augustin_ciarlet1988}.
Additionally,
\begin{equation}
 \symbtensor{\varepsilon}(\vec{v}) =
 \frac{1}{2}\left(\operatorname{grad}\vec{v}+ ( \operatorname{grad}
 \vec{v})^\top \right)
\end{equation}
and $\mathcal{A}_t$ is the
nonlinear operator in the current configuration which is induced
by the stress tensor representation
\eqref{Augustin_eq:additiveStress}, and by using the related
duality pairing $\langle \cdot , \cdot \rangle_{\Omega_t}$. For
later use, we introduce the corresponding terms in the reference
configuration $\Omega_0$ as $\langle \mathcal{A}_0(\vec{U}),
\vec{V}\rangle_{\Omega_0}$ and $\langle
\vec{\mathcal{F}}_0,\vec{V}\rangle_{\Omega_0}$. Note that
\eqref{Augustin_eq:WeakCurrent} formally corresponds to a
variational formulation in linear elasticity. However, the
integral and the involved terms have to be evaluated in the
current configuration which comprises the nonlinearity of the
system. If the test function $\vec{v}$ is interpreted as the
spatial velocity gradient, then
$\symbtensor{\varepsilon}(\vec{v})$ is the rate of deformation
tensor so that $\langle \mathcal{A}_t(\vec{u}), \vec{v}
\rangle_{\Omega_t}$ has the physical interpretation of the rate of
internal mechanical work.

In terms of the reference configuration, the boundary value problem \eqref{Augustin_eq:reference}, \eqref{Augustin_eq:Neumann}
is formally equivalent to the variational equations
\begin{equation}
\langle \mathcal{A}_0(\vec{U}),\vec{V} \rangle_{\Omega_0} =
\int_{\Omega_0} \tensor{S}(\vec{U}) : {\symbtensor{\Sigma}}(\vec{U},\vec{V})
\dd \vec{X} = \int_{\Omega_0} \vec{b}_0 \cdot \vec{V} \dd \vec{X} +
\int_{\Gamma_{0,\mathrm{N}}} \vec{G}_0 \cdot \vec{V} \dd s_{\vec{X}} =
\langle \vec{\mathcal{F}}_0,\vec{V} \rangle_{\Omega_0},
\label{Augustin_eq:WeakReference}
\end{equation}
valid for a smooth enough tensor field $\tensor{S}(\vec{U}):\overline{\Omega}_0\mapsto\mathbb{R}^{3\times 3}$
and all smooth enough vector fields $\vec{V}:\overline{\Omega}_0\mapsto\mathbb{R}^3$
with $\vec{V}=\vec{0}$ on $\Gamma_{0,\mathrm{D}}$, see, e.g., \cite[Theorem~2.6-1]{Augustin_ciarlet1988}.
In \eqref{Augustin_eq:WeakReference} we use the definition of
the directional derivative of the Green--Lagrange strain tensor, i.e.
\begin{equation}
\symbtensor{\Sigma}(\vec{U},\vec{V}) =
\frac{1}{2}\left(\operatorname{Grad}^\top\vec{V}\,\tensor{F}(\vec{U})+\tensor{F}^\top
(\vec{U})\operatorname{Grad}\vec{V}\right),
\end{equation}
which is also known as the variation or the material time
derivative of the Green--Lagrange strain tensor in the literature.

It is important to note that results on existence of solutions in
nonlinear elasticity can be stated given a polyconvex
strain-energy function $\Psi$, which holds true for the anisotropic model
\eqref{Augustin_eq:isoaniso} discussed in Section~\ref{Augustin_sec:model}.
For more details we refer to the
results of Ball~\cite{Augustin_ball1977,Augustin_ball1976}, see
also \cite{Augustin_ciarlet1988, Augustin_dacorogna2008} and
Balzani et al. \cite{Augustin_balzani2006}.
\subsection{Linearization and discretization}
In the following we confine ourselves to the reference configuration
$\Omega_0$. The formulations in the current configuration $\Omega_t$
can be deduced in an analogous way.

For the solution of the nonlinear system
\eqref{Augustin_eq:WeakReference} we apply Newton's method to obtain
the recursion
\begin{equation}\label{Augustin_eq:NewtonRecursion}
\langle \Delta\vec{U}, \mathcal{A}_0'(\vec{U}^k) \vec{V}
\rangle_{\Omega_0} = \langle \vec{\mathcal{F}}_0,\vec{V} \rangle_{\Omega_0} -
\langle \mathcal{A}_0(\vec{U}^k),\vec{V} \rangle_{\Omega_0}, \quad
\vec{U}^{k+1}=\vec{U}^k+\Delta\vec{U},
\end{equation}
with the tangential term $\mathcal{A}_0'(\vec{U}^k)$, the displacement field of the $k$-th Newton step $\vec{U}^k$, the
increment $\Delta\vec{U}$ and a suitable initial guess $\vec{U}^0$.

For the computational domain $\Omega_0 \subset \mathbb{R}^3$ we consider
an admissible decomposition into $N$ tetrahedral shape regular finite
elements $\tau_\ell$ of mesh size $h_\ell$, i.e.
$\overline{\Omega}_0=\overline{\mathcal{T}}_N=\bigcup_{\ell=1}^N
\overline{\tau}_\ell$, and we introduce a conformal finite element
space $X_h \subset [H^1(\Omega_0)]^3$, $M = \text{dim} X_h$,
of piecewise polynomial continuous
basis functions $\varphi_i$. Then the Galerkin finite element
discretization of the linearized variational formulation
\eqref{Augustin_eq:NewtonRecursion} results in a system of
algebraic equations to find $\Delta\vec{U}_h \in X_h$, $\Delta\vec{U}_h=\vec{0}$ on  $\Gamma_{0,\mathrm{D}}$ such that
\begin{equation}\label{Augustin_eq:NewtonRecursion_h}
\langle \Delta\vec{U}_h, \mathcal{A}_0'(\vec{U}_h^k) \vec{V}_h
\rangle_{\Omega_0} = \langle \vec{\mathcal{F}}_0,\vec{V}_h \rangle_{\Omega_0} -
\langle \mathcal{A}_0(\vec{U}_h^k),\vec{V}_h \rangle_{\Omega_0}, \quad
\vec{U}^{k+1}_h=\vec{U}_h^k+\Delta\vec{U}_h,
\end{equation}
holds for all $\vec{V}_h \in X_h$, $\vec{V}_h = \vec{0}$ on $\Gamma_{0,\mathrm{D}}$.
Note that the initial guess $\vec{U}^0_h$ has to satisfy an
approximate Dirichlet boundary condition $\vec{U}^0_h = \vec{U}_{\mathrm{D},h}$
on $\Gamma_{0,\mathrm{D}}$ to fulfill condition \eqref{Augustin_eq:Dirichlet}, where
$\vec{U}_{\mathrm{D},h} \in X_{h|\Gamma_0,\mathrm{D}}$ denotes a suitable
approximation of the given displacement $\vec{U}_{\mathrm{D}}$.
For the computation of the tangential term
$\mathcal{A}_0'(\vec{U}_h^k)$ we need to evaluate
\begin{align}
\langle \Delta\vec{U}_h, \mathcal{A}_0'(\vec{U}_h^k)\vec{V}_h \rangle_{\Omega_0}
&= \int_{\Omega_0} \operatorname{Grad}(\Delta\vec{U}_h) \,
\tensor{S} (\vec{U}_h^k) : \operatorname{Grad} \vec{V}_h \dd \vec{X} \nonumber\\
 & \quad +
\int_{\Omega_0} \tensor{F}^\top \operatorname{Grad}{\Delta\vec{U}_h} :
\mathbb{C}(\vec{U}_h^k) : \tensor{F}^\top \operatorname{Grad}(\vec{V}_h) \dd \vec{X}.
  \label{Augustin_eq:Tangent}
\end{align}
For a more detailed presentation how to compute the tangential
term, in particular the forth-order elasticity tensor
$\mathbb{C}(\vec{U}_h^k)$ we refer to
\cite{Augustin_holzapfel2003,Augustin_augustindiss2012}.

Note that the convergence rate of the Newton method is dependent on the
initial guess, on the parameters used in the model and on the inhomogeneous
Dirichlet and Neumann boundary conditions which influence
$\vec{\mathcal{F}}_0$.

In a time-stepping scheme we use zero for the initial guess, and
the result of the $k$-th time step as initial solution for the
next step. The initial guess may also be the solution of a
modified nonlinear elasticity problem such as the solution of the
same nonlinear model but with modified parameters, e.g., a reduced
penalty parameter $\kappa$, or modified boundary conditions, e.g.,
a reduced pressure on the surface. The latter is equivalent to an
incremental load stepping scheme with a parameter $\tau\in(0,1],\
\tau\to 1$, so that
\begin{equation}\label{Augustin_eq:loadStepping}
\langle \Delta\vec{U}_h,A'(\vec{U}_h^k)\vec{V}_h\rangle_{\Omega_0} =
\langle \tau\vec{\mathcal{F}}_0,\vec{V}_h \rangle_{\Omega_0} -
\langle A(\vec{U}_h^k),\vec{V}_h\rangle_{\Omega_0},\quad
\vec{U}^{k+1}_h=\vec{U}_h^k+\Delta\vec{U}_h.
\end{equation}
Klawonn and Rheinbach \cite{Augustin_klawonn2010}
used a load stepping scheme of this kind, for more information on
load stepping and global Newton methods, see
\cite{Augustin_wriggers2008,Augustin_deuflhard2011}. The standard
finite element method (FEM) now yields a linear system of
equations which is equivalent to the discretized variational
formulation \eqref{Augustin_eq:NewtonRecursion_h}. Finally, we have
to solve
\begin{equation}\label{Augustin_eq:LGS}
\tensor{K}'(\Rvec{U}^k)\,\Delta\Rvec{U} =
\Rvec{\mathcal{F}}-\Rvec{K}(\Rvec{U}^k),\quad
\Rvec{U}^{k+1}=\Rvec{U}^k+\Delta\Rvec{U},
\end{equation}
with the solution vector $\Rvec{U}^k$ in the $k$-th Newton step and the
increment $\Delta\Rvec{U}$. The tangent stiffness matrix $\tensor{K}'$
is calculated according to
\begin{equation}
\tensor{K}'(\Rvec{U}^k)[i,j] := \langle
\vec{\varphi}_j,A'(\vec{U}_h^k)\,\vec{\varphi}_i
\rangle_{\Omega_0},
\end{equation}
and the terms of the right hand side are constructed by
\begin{equation}
\Rvec{\mathcal{F}}[i] := \langle
\vec{\mathcal{F}}_0,\vec{\varphi}_i\rangle_{\Omega_0} \quad \text{
and } \quad  \Rvec{K}(\Rvec{U}^k)[i] := \langle
A(\vec{U}_h^k),\vec{\varphi}_i \rangle_{\Omega_0}.
\end{equation}
The additive split of the stress tensors
\eqref{Augustin_eq:additiveStress} and the introduction of the
hydrostatic pressure \eqref{Augustin_eq:hydrostaticPressure} leads
to the additional equation
\begin{equation}\label{Augustin_eq:volumetricVariationDis}
p-\frac{\partial\Psi_\text{vol}(J)}{\partial J}=0,
\end{equation}
which has to be satisfied in a weak sense. For this we use the
idea of \textit{static condensation} where this volumetric
variable is eliminated element-wise, see, e.g.,
\cite{Augustin_holzapfel2003}. This may be achieved in using
discontinuous basis functions; in this paper we will concentrate
on piecewise constants. In the case of tetrahedral elements, this
approach leads to $\mathcal{P}_k-\mathcal{P}_0$ elements. Here
$k$ is the order of the basis functions for the displacement
field. It is known that linear finite elements are very prone to
volumetric locking. Hence, for nearly incompressible materials
piecewise quadratic elements ($k=2$) are a better choice,
see Simo \cite{Augustin_simo1998}. The resulting
$P_2-P_0$ element is also the preferred choice to
model nearly incompressible arterial materials in
\cite{Augustin_klawonn2010}--\cite{Augustin_balzani2012}.
For the numerical results in this work
(Section~\ref{Augustin_sec:numericalResults}) we use both linear
($P_1-P_0$ element) and quadratic ($P_2-P_0$ element)
ansatz functions for the displacement field and compare the
results.

Note that due to the symmetry of the stress tensor $\tensor{S}$
and the major and minor symmetry properties of the elasticity
tensor $\mathbb{C}$ the operator $A_0'(\vec{U}^k)$ is self-adjoint.
We can also show, using the positive definiteness of the
elasticity tensor, see \cite{Augustin_ogden1997}, and the
polyconvexity of the strain-energy function
(Section~\ref{Augustin_sec:varForm}), that this operator is
$[H_0^1(\Omega_0,\Gamma_{0,\mathrm{D}})]^3$-elliptic and bounded,
see \cite{Augustin_ogden1997,Augustin_augustindiss2012}. With
these properties of the operator $A_0'(\vec{U}_h^k)$ we can state
that the linearized system
\eqref{Augustin_eq:NewtonRecursion_h},\eqref{Augustin_eq:Tangent}
admits a unique solution $\Delta \vec{U}_h$. Furthermore, the
tangent stiffness matrix $\tensor{K}'$ is symmetric and positive
definite.

Simulations with large deformations and the hence required
derivative of the Neumann boundary conditions
\eqref{Augustin_eq:Neumann} would yield an additional
non-symmetric mass matrix on the left hand side of
\eqref{Augustin_eq:LGS}. To stay with an symmetric system we
neglect this matrix but compensate it with a surface update of the
geometry after each Newton step. Thus, our whole system is
symmetric and we can use the CG method as an iterative solver.
Nonetheless, the FETI methods described in
Section~\ref{Augustin_sec:FETI} also work for non-symmetric systems by using the GMRES method.

\section{Finite Element Tearing and Interconnecting}
\label{Augustin_sec:FETI} To solve the linearized equations
\eqref{Augustin_eq:LGS} arising in the Newton method we apply the
finite element tearing and interconnecting approach
\cite{Augustin_farhat1991}, see also
\cite{Augustin_klawonn2010,Augustin_pechstein2013,Augustin_toselli2005},
and references given therein. The derivation of the FETI system
for nonlinear mechanics will be performed in the reference
configuration. In an analogous way this is also valid for the
formulation in the current configuration.
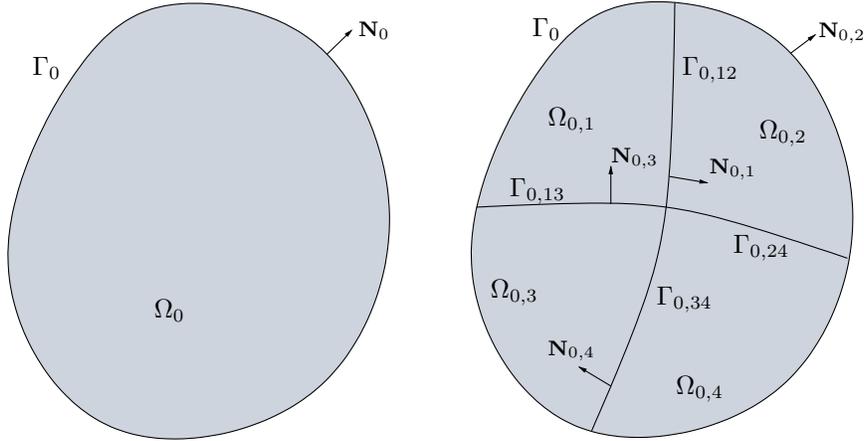
\begin{figure}[b!]
  \centering
  $
  \vcenter{\hbox{\input{augustin_domain.tex}}}\qquad\quad
  \vcenter{\hbox{\input{augustin_decomposed.tex}}}
  $
  \caption{Decomposition of a domain $\Omega_0$ into four subdomains $\Omega_{0,i}$, $i=1,\ldots,4$.}
  \label{Augustin_fig:decomposition}
\end{figure}
For a bounded domain $\Omega_0 \subset \mathbb{R}^3$ we introduce a
non-overlapping domain decomposition
\begin{equation} \label{Augustin_eq:decomposition}
  \overline{\Omega}_0 = \bigcup\limits_{i=1}^p \overline{\Omega}_{0,i} \quad
  \text{with} \; \Omega_{0,i} \cap \Omega_{0,j} = \emptyset \quad
  \text{for} \; i \neq j,
  \quad \Gamma_{0,i} = \partial \Omega_{0,i},
\end{equation}
see Fig.~\ref{Augustin_fig:decomposition}.
The local interfaces are given by $\Gamma_{0,ij}:=\Gamma_{0,i}\cap\Gamma_{0,j}$
for all $i<j$. The skeleton of the domain decomposition
(\ref{Augustin_eq:decomposition}) is denoted as
\begin{equation} \label{Augustin_eq:skeleton}
  \Gamma_{0,\mathrm{C}} := \bigcup_{i=1}^p \Gamma_{0,i} =
  \Gamma_0 \cup \bigcup_{i<j} \overline{\Gamma}_{0,ij} .
\end{equation}
We assume that the finite element mesh $\mathcal{T}_N$ matches
the domain decomposition \eqref{Augustin_eq:decomposition}, i.e.,
we can reorder the degrees of freedom to rewrite the linear system
(\ref{Augustin_eq:LGS}) as
\begin{equation}\label{Augustin_eq:reordered}
  \begin{aligned}
\left( \begin{array}{cccc}
  \tensor{K}_{11}'(\Rvec{U}_{1}^k) & & &
  \tensor{K}_{1\mathrm{C}}'(\Rvec{U}_{1}^k) \tensor{A}_1 \\[1mm]
& \ddots & & \vdots \\[1mm]
& & \tensor{K}_{pp}'(\Rvec{U}_{p}^k) &
 \tensor{K}_{p\mathrm{C}}'(\Rvec{U}_{p}^k) \tensor{A}_p \\[1mm]
\tensor{A}_1^\top \tensor{K}_{\mathrm{C}1}'(\Rvec{U}_{1}^k) & \cdots &
\tensor{A}_p^\top \tensor{K}_{\mathrm{C}p}'(\Rvec{U}_{p}^k) &
 \displaystyle\sum\limits_{i=1}^p \tensor{A}_i^\top \tensor{K}_{\mathrm{CC}}'(\Rvec{U}_{i}^k)
\tensor{A}_i
\end{array} \right)
\left( \begin{array}{c} \Delta \Rvec{U}_{1,I}^k \\[1mm] \vdots \\[1mm]
  \Delta \Rvec{U}_{p,I}^k \\[1mm] \Delta \Rvec{U}_\mathrm{C}^k \end{array}
\right) \\
 = - \left( \begin{array}{c}
  \Rvec{K}_1(\Rvec{U}_{1}^k) \\[1mm] \vdots \\[1mm] \Rvec{K}_p(\Rvec{U}_{p}^k) \\[1mm]
  \sum\limits_{i=1}^p \tensor{A}_i^\top \Rvec{K}_\mathrm{C}(\Rvec{U}_{i}^k)
\end{array} \right), &&
\end{aligned}
\end{equation}
where the increments $\Delta \Rvec{U}_{i,I}^k$, the stiffness matrices
$\tensor{K}_{ii}'(\Rvec{U}_{i}^k)$ and the terms on the right hand side
$\Rvec{K}_i(\Rvec{U}_{i}^k),\ i=1,\dots,p$, are related to the local degrees
of freedom within the subdomain $\Omega_{0,i}$. All terms with an index
$\mathrm{C}$
correspond to degrees of freedom on the coupling boundary
$\Gamma_{0,\mathrm{C}}$, see
\eqref{Augustin_eq:skeleton}, while $\tensor{A}_i$ denote simple
reordering matrices taking boolean values.

\subsection{Classical FETI method}\label{Augustin_sec:classicalFeti}
Starting from \eqref{Augustin_eq:reordered}, the \textit{tearing} is now
carried out by
\begin{equation}
\Delta\Rvec{U}_i = \left( \begin{array}{c}
\Delta\Rvec{U}_{i,I}^k \\[1mm]
\tensor{A}_i \Delta\Rvec{U}_\mathrm{C}^k
\end{array} \right), \;
\tensor{K}_i' = \left( \begin{array}{cc}
  \tensor{K}_{ii}'(\Rvec{U}_{i}^k) & \tensor{K}_{i\mathrm{C}}'(\Rvec{U}_{i}^k) \\[1mm]
  \tensor{K}_{\mathrm{C}i}'(\Rvec{U}_{i}^k) & \tensor{K}_{\mathrm{CC}}'(\Rvec{U}_{i}^k)
\end{array} \right), \;
\Rvec{f}_i = - \left( \begin{array}{c}
  \Rvec{K}_i(\Rvec{U}_{i}^k) \\[1mm] \Rvec{K}_\mathrm{C}(\Rvec{U}_{i}^k)
\end{array} \right),
\end{equation}
where $\tensor{A}_i \Delta\Rvec{U}_\mathrm{C}^k$ is related to
degrees of freedom on the coupling boundary $\Gamma_{0,i}
\backslash \Gamma_0$. As the unknowns $\Delta\Rvec{U}_i$ are
typically not continuous over the interfaces we have to ensure the
continuity of the solution on the interface, i.e.
\begin{equation}\label{Augustin_eq:jumpConstraints}
  \Delta\Rvec{U}_{i}=\Delta\Rvec{U}_{j}\quad \text{on}\ \Gamma_{0,ij},\
  i,j=1,\dots,p.
\end{equation}
This is done by applying the \textit{interconnecting}
\begin{equation}\label{Augustin_eq:jumpConstraints 1}
  \sum\limits_{i=1}^p \tensor{B}_i \Delta\Rvec{U}_i = \Rvec{0},
\end{equation}
where the matrices $\tensor{B}_i$ are constructed from $\{0,1,-1\}$ such that
\eqref{Augustin_eq:jumpConstraints} holds. By using
discrete Lagrange multipliers $\Rvec{\lambda}$ to enforce the constraint
\eqref{Augustin_eq:jumpConstraints 1} we finally have to solve the
linear system
\begin{equation}\label{Augustin_eq:classicalFetiLGS}
\left( \begin{array}{cccc}
\tensor{K}_1' & & & \tensor{B}_1^\top \\
 & \ddots & & \vdots \\
 & & \tensor{K}_p' & \tensor{B}_p^\top \\
 \tensor{B}_1 & \ldots & \tensor{B}_p & \tensor{0}
\end{array} \right)
\left( \begin{array}{c} \Delta\Rvec{U}_1 \\ \vdots \\ \Delta\Rvec{U}_p \\
\Rvec{\lambda} \end{array} \right) =
\left( \begin{array}{c}
\Rvec{f}_1 \\ \vdots \\ \Rvec{f}_p \\ \Rvec{0}
\end{array} \right) .
\end{equation}

\subsection{all-floating FETI method}\label{Augustin_sec:allfloating}
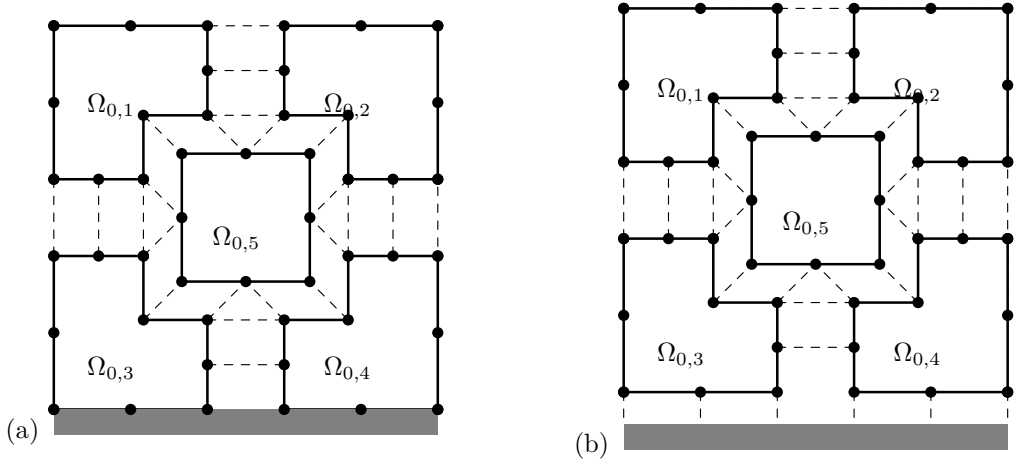
\begin{figure}[!t]
  \centering
\begin{minipage}[c]{0.5\linewidth}(a)
      \def\svgwidth{60mm}\input{augustin_classical.tex}
\end{minipage}\hfill
\begin{minipage}[c]{0.5\linewidth}(b)
      \def\svgwidth{60mm}\input{augustin_allfloating.tex}
\end{minipage}
\caption{Fully redundant classical FETI (a) and all-floating FETI
(b)
  formulation: $\Omega_{0,i}$, $i=1,\dots,5$, denote the local subdomains,
  the black dots correspond to the subdomain
  vertices and the dashed lines correspond to the constraints
  \eqref{Augustin_eq:jumpConstraints}. The gray strip indicates Dirichlet
  boundary conditions. Note that the number of constraints for the all-floating
  approach rises with the number of vertices on the Dirichlet boundary.}
  \label{Augustin_fig:allfloating}
  \end{figure}

The idea of this special FETI method, cf., e.g., Of and Steinbach
\cite{Augustin_of2009}, is to treat {all} subdomains as floating
subdomains, i.e. domains with no Dirichlet boundary conditions. In
addition to the standard procedure of `gluing' the subregions
along the auxiliary interfaces, the Lagrange multipliers are now
also used for the implementation of the Dirichlet boundary
conditions, see Fig.~\ref{Augustin_fig:allfloating}. This
simplifies the implementation of the FETI procedure since it is
possible to treat all subdomains in the same way. In addition,
some tests (Section~\ref{Augustin_sec:numericalResults}) show more
efficiency than the classical FETI approach and the asymptotic
behavior improves. This is due to the mapping properties of the
Steklov--Poincar\'e operator, see \cite[Remark
1]{Augustin_of2009}. The drawback is an increasing number of
degrees of freedom and Lagrange multipliers. Compare also to
Dost{\'a}l et al. \cite{Augustin_dostal2006} for the related
Total-FETI method.
If all regions are treated as floating subdomains the conformance of the Dirichlet boundary conditions is not given;
 they have to be enhanced in the system of constraints using the slightly
modified interconnecting
\begin{equation} \label{Augustin_eq:jumpoperatorAF}
  \sum\limits_{i=1}^p \tilde{\tensor{B}}_i\Delta\Rvec{U}_i=\Rvec{b},
\end{equation}
where $\tilde{\tensor{B}}_i$ is a block matrix of the kind
$\tilde{\tensor{B}}_i=[ \tensor{B}_i,\tensor{B}_{\mathrm{D},i}]^\top$ and the vector
$\Rvec{b}$ is of the form $\Rvec{b}=[\Rvec{0},\Rvec{b}_\mathrm{D}]^\top$ such that
$\tensor{B}_{\mathrm{D},i}[j,k]=1$, if and only if $k$
is the index of a Dirichlet node $j$ of the subdomain $\Omega_i$,
while $\Rvec{b}[j]$ equals the Dirichlet values
corresponding to the vertices $\vec{X}_k\in\Gamma_{0,\mathrm{D}}$, see also
\cite{Augustin_of2009}.

For three-dimensional elasticity problems all subdomain stiffness matrices
have now the same and known defect, which equals the number of six rigid body
motions and which also
simplifies the calculation of the later needed generalized
inverse matrices $\tensor{K}_i^\dagger$.
For all-floating FETI we finally get the linearized system of
equations
\begin{equation}\label{Augustin_eq:AFFETI}
\left( \begin{array}{cccc}
  \tensor{K}_1' & & & \tilde{\tensor{B}}_1^\top \\
 & \ddots & & \vdots \\
 & & \tensor{K}_p' & \tilde{\tensor{B}}_p^\top \\
 \tilde{\tensor{B}}_1 & \ldots & \tilde{\tensor{B}}_p & \tensor{0}
\end{array} \right)
\left( \begin{array}{c} \Delta\Rvec{ U}_1 \\ \vdots \\ \Delta\Rvec{U}_p \\
\Rvec{\lambda} \end{array} \right) =
\left( \begin{array}{c}
\Rvec{f}_1 \\ \vdots \\ \Rvec{f}_p \\ \Rvec{b}
\end{array} \right) .
\end{equation}

\subsection{Solving the FETI system}\label{Augustin_sec:solving}
To solve the linearized systems
\eqref{Augustin_eq:classicalFetiLGS} and
\eqref{Augustin_eq:AFFETI} we follow the standard approach of
tearing and interconnecting methods. For convenience we outline
the procedure by means of the classical FETI formulation
(Section~\ref{Augustin_sec:classicalFeti}). However the {\em modus
operandi} is analogous for the all-floating approach.

First, note that in the case of a floating subdomain $\Omega_{0,i}$, i.e.
$\Gamma_{0,i} \cap \Gamma_{0,\mathrm{D}} = \emptyset$, the local matrices
$\tensor{K}_i'$ are not
invertible. Hence, we introduce a generalized inverse $\tensor{K}_i^\dagger$
to represent the local solutions as
\begin{equation}\label{Augustin_eq:localSolutions}
 \Delta\Rvec{U}_i = \tensor{K}_i^\dagger
 (\Rvec{f}_i - \tensor{B}_i^\top \Rvec{\lambda})
 + \sum\limits_{k=1}^6 \gamma_{k,i} \Rvec{r}_{k,i}.
\end{equation}
Here, $\Rvec{r}_{k,i} \in \operatorname{ker} \, \tensor{K}_i'$ correspond to the
rigid body motions of elasticity and $\gamma_{k,i}$ are unknown constants.
For floating subdomains we additionally
require the solvability conditions
\begin{equation}
 (\Rvec{f}_i - \tensor{B}_i^\top \Rvec{\lambda} , \Rvec{r}_{k,i} ) = 0
 \quad \mbox{for} \quad i=1,\ldots,6.
\end{equation}
In the case of a non-floating subdomain, i.e. $\operatorname{ker}
\, \tensor{K}'_i = \emptyset$, we may set $\tensor{K}_i^\dagger =
\tensor{K}_i^{-1}$. Note that it may happen that the
kernel $\operatorname{ker} \, \tensor{K}'_i$ is non-trivial and
its dimension is lower than 6. This is the case if the set
$\Gamma_{0,i} \cap \Gamma_{0,\mathrm{D}}$ is either a vertex or an
edge. For classical FETI methods this requires the implementation
of an effective method to identify these kernels reliably. Note
that this is a key advantage of the all-floating FETI approach
because all subdomains are here treated as floating subdomains,
and hence we know the kernel of each local operator
$\operatorname{ker} \, \tensor{K}'_i=6$. With these kernels the
solution of the local problems to find the generalized inverse
$\tensor{K}_i^\dagger$ can be reduced to sparse systems which are
typically better conditioned as the systems arising from the
FETI-DP method, see Brzobohat{\`y} et al.
\cite{Augustin_brzobohaty2011}. In
Section~\ref{Augustin_sec:allfloating} we comment on an
all-floating approach where also Dirichlet boundary conditions
are incorporated by using discrete Lagrange multipliers.

In general, the Schur complement system of \eqref{Augustin_eq:classicalFetiLGS}
is constructed to obtain
\begin{equation}
\sum\limits_{i=1}^p \tensor{B}_i \tensor{K}_i^\dagger
\tensor{B}_i^\top \Rvec{\lambda} - \sum\limits_{i=1}^p
\sum\limits_{k=1}^6 \gamma_{k,i} \tensor{B}_i \Rvec{r}_{k,i} =
\sum\limits_{i=1}^p \tensor{B}_i \tensor{K}_i^\dagger \Rvec{f}_i ,
\quad ( \Rvec{f}_i - \tensor{B}_i^\top \Rvec{\lambda} ,
\Rvec{r}_{k,i} ) = 0.
\end{equation}
This can be expressed as
\begin{equation}\label{Augustin_eq:classicalFetiF}
\left( \begin{array}{cc} \tensor{F}\phantom{^\top} & - \tensor{G} \\[1mm]
\tensor{G}^\top &
  \phantom{-}\tensor{0}\end{array} \right)
\left( \begin{array}{c} \Rvec{\lambda} \\[1mm] \Rvec{\gamma}
\end{array} \right) = \left( \begin{array}{c} \Rvec{d} \\[1mm]
\Rvec{e} \end{array} \right),
\end{equation}
with
\begin{equation}
\tensor{F} = \sum\limits_{i=1}^p \tensor{B}_i \tensor{K}_i^\dagger
\tensor{B}_i^\top, \; \tensor{G} = \sum\limits_{i=1}^p
\sum\limits_{k=1}^6 \tensor{B}_i \Rvec{r}_{k,i}, \; \Rvec{d} =
\sum\limits_{i=1}^p \tensor{B}_i \tensor{K}_i^\dagger \Rvec{f}_i,
\end{equation}
and $\Rvec{e}$ is constructed using $e_{k,i} =
(\Rvec{f}_i,\Rvec{r}_{k,i})$ for $i=1,\dots,p$ and $k=1,\dots,6$.
For the solution of the linearized system
\eqref{Augustin_eq:classicalFetiF} the projection
\begin{equation}\label{Augustin_eq:projection}
\tensor{P}^\top := \tensor{I} - \tensor{G} \left( \tensor{G} \tensor{G}^\top \right)^{-1} \tensor{G}^\top
\end{equation}
is introduced. It now remains to consider the projected system
\begin{equation}\label{Augustin_eq:dualProblem}
\tensor{P}^\top \tensor{F} \Rvec{\lambda} = \tensor{P}^\top \Rvec{d}.
\end{equation}
This can be solved by using a parallel iterative method
with suitable preconditioning of the form
\begin{equation}\label{Augustin_eq:feti_preconditioner}
  \tensor{M}^{-1} := \sum\limits_{i=1}^p \tensor{B}_{\mathrm{D},i}
  \tensor{Y}_i \tensor{B}_{\mathrm{D},i}^\top,
\end{equation}
with modified jump operators $\tensor{B}_{\mathrm{D},i}$ which are
obtained by multiplicity scaling, see
\cite{Augustin_klawonn2010,Augustin_toselli2005}.
Since the local subproblems all yield symmetric tangent stiffness
matrices $\tensor{K}'_i,\ i=1,\dots,p$,
cf.~Section~\ref{Augustin_sec:FEM}, the matrix
$\tensor{P}^\top\tensor{F}$ is also symmetric. This enables us to
use the CG method as the global solver for
\eqref{Augustin_eq:dualProblem}. Be aware that the initial
approximate solution $\Rvec{\lambda}^0$ has to satisfy the
compatibility condition $\tensor{G}^\top \Rvec{\lambda}^0 =
\Rvec{e}$. A possible choice is
\begin{equation}
\Rvec{\lambda}_0=\tensor{G}\left(\tensor{G}^\top\tensor{G}\right)^{-1}\Rvec{e}.
\end{equation}
In a post processing we finally recover the vector of constants
\begin{equation}
 \Rvec{\gamma} =\left( \tensor{G}^\top \tensor{G} \right)^{-1}
 \tensor{G}^\top \left( \tensor{F} \Rvec{\lambda}-\Rvec{d} \right),
\end{equation}
and subsequently the desired solution
\eqref{Augustin_eq:localSolutions}.
\subsection{Preconditioning}\label{Augustin_sec:preconditioning}
Following Farhat et al. \cite{Augustin_farhat1994a} we apply
either the lumped preconditioner
\begin{equation}\label{Augustin_eq:LumpedVK}
  \tensor{M}^{-1}_\mathrm{L} := \sum\limits_{i=1}^p \tensor{B}_{\mathrm{D},i} \tensor{K}_i' \tensor{B}_{\mathrm{D},i}^\top,
\end{equation}
or the optimal Dirichlet preconditioner
\begin{equation}\label{Augustin_eq:DirichletVK}
  \tensor{M}^{-1}_\mathrm{D} := \sum\limits_{i=1}^p \tensor{B}_{\mathrm{D},i}  \left( \begin{array}{cc}
\tensor{0} & \tensor{0} \\[1mm] \tensor{0} & \tensor{S}_i \end{array} \right) \tensor{B}_{\mathrm{D},i} ^\top,
\end{equation}
where
\begin{equation}
  \tensor{S}_i = \tensor{K}_{\mathrm{CC}}'(\Rvec{U}_{i}^k) -
  \tensor{K}_{\mathrm{C}i}'(\Rvec{U}_{i}^k) \tensor{K}_{ii}'^{-1}(\Rvec{U}_{i}^k)
  \tensor{K}_{i\mathrm{C}}'(\Rvec{U}_{i}^k)
\end{equation}
is the Schur complement of the local finite element matrix
$\tensor{K}_i'$. Alternatively, one may also use scaled
hypersingular boundary integral operator preconditioners, as
proposed in \cite{Augustin_langer2003}.
For comparison we employ an identity preconditioner
which is constructed by using the identity matrix for
$\tensor{Y}_i$ in eq.~(\ref{Augustin_eq:feti_preconditioner}).

\section{Numerical Results}\label{Augustin_sec:numericalResults}
In this section some representative numerical
examples for the finite element tearing and interconnecting
approach for linear and nonlinear elasticity problems are
presented. First, the FETI implementation is tested within linear
elasticity. Here we are able to compare the computed results to a
given exact solution. This enables us to show the efficiency of
our implementation and also the convergence rates, as predicted
from the theory. We compare the different preconditioning
techniques and present differences between the classical FETI and
the all-floating FETI approach.

Subsequently, we apply the FETI method to nonlinear elasticity
problems. Thereby, we focus on the anisotropic model, as described
in Section~\ref{Augustin_sec:model}, and use a realistic
triangulations of the aorta and a common carotid artery. As in the
linear elastic case, different preconditioning techniques for the
all-floating and for the classical FETI method are compared.
In Section~\ref{sec:loadStepping}, we analyze the
biomechanical behavior of an aorta up to an internal pressure of
$300$\,mmHg and plot stress and displacement evolutions as a
function of the internal pressure. Finally, in
Section~\ref{strongscaling}, we analyze our computational
framework with respect to strong scaling properties.

The calculations were performed by using the \textit{VSC2}-cluster
(\textsf{http://vsc.ac.at/}) in Vienna. This Linux cluster
features 1314 compute nodes, each with two AMD Opteron Magny Cours
6132HE (8 Cores, 2.2 GHz) processors and 8 x 4 RAM. This yields
the total number of $21\,024$ available processing units. As local
direct solver we use Pardiso
\cite{Augustin_schenk2008,Augustin_schenk2007}, included in
Intel's Math Kernel Library (MKL).

\subsection{Linear elasticity}\label{Augustin_sec:NRLinElast}
In this section of numerical benchmarks we consider a linear
elastic problem with the academic example of a unit cube which is
decomposed into a certain number of subcubes. Dirichlet boundary
conditions are imposed all over the surface
$\Gamma_\mathrm{D}=\partial\Omega$. The parameters used are
Young's modulus $E=210$ GPa and Poisson's ratio $\nu=0.45$. The
calculated solution is compared to the fundamental solution of
linear elastostatics
\begin{equation}\label{Augustin_eq:fundSol}
  {U}_{1k}^\ast(\vec{x},\vec{x}^\ast)=\frac{1}{8\pi}\frac{1}{E}\frac{1+\nu}{1-\nu}\left[(3-4\nu)\frac{\delta_{1k}}{|\vec{x}-\vec{x}^\ast|}+\frac{(x_1-x^\ast_1)(x_l-x^\ast_l)}{|\vec{x}-\vec{x}^\ast|^3}\right],\ k=1,2,3
\end{equation}
for all $\vec{x}\in\Omega$, $\vec{x}^\ast\in\mathbb{R}^3$ is an
arbitrary point outside of the domain $\Omega$, and $\delta_{ij}$
is the Kronecker delta, see
\cite{Augustin_steinbach2008}. The different strategies of
preconditioning are compared and also the all-floating and
classical FETI approaches. As global iterative method we use the
CG method with a relative error reduction of
$\varepsilon=10^{-8}$. Under consideration is a linear elasticity
problem using linear tetrahedral elements
($\mathcal{P}_1$ elements) with a uniform refinement over five
levels ($\ell = 1,\dots, 5$) given a cube with $512$ subdomains.

Hence, the number of degrees of freedom associated with the
coarsest mesh is $9\,981$ for the all-floating FETI approach and
$6\,621$ for the classical FETI approach. The difference of the
numbers is due to the decoupling of the Dirichlet boundary
$\Gamma_\mathrm{D}$. For the finest mesh we have $31\,116\,861$
(all-floating) and $31\,073\,181$ (classical) degrees of freedom.
The number of Lagrange multipliers varies between $38\,052$ for
level $1$ and $2\,908\,692$ for level 5. Again we have a higher
number of Lagrange multipliers for the all-floating approach due
to the decoupling of the Dirichlet boundary conditions. The
computations were performed on VSC2 using $512$ processing units.

First note in Table~\ref{Augustin_tab:cube512_linear} that for
all examined settings, the L2 error, i.e.
\begin{equation}
  \lVert \vec{u}-\vec{u}_h\rVert_{L_2(\Omega)},
  \label{eq:L2Error}
\end{equation}
where $\vec{u}_h$ is the approximate and $\vec{u}$ the exact solution,
and the estimated order of convergence
\begin{equation}
  \mathrm{eoc}_\ell = \frac{\ln\lVert
    \vec{u}-\vec{u}_{h,\ell}\rVert_{L_2(\Omega)}-\ln\lVert
  \vec{u}-\vec{u}_{h,\ell+1}\rVert_{L_2(\Omega)}}{\ln 2}
  \label{eq:L2Errora}
\end{equation}
behaves as predicted from the theory, i.e. it is of second order.
As expected the least iteration numbers were observed for the
optimal Dirichlet preconditioner. Nonetheless, since no additional
time is required to compute the lumped preconditioner, in contrast
to the more sophisticated Dirichlet preconditioner, this type of
preconditioning yields comparable computational times for each
level of refinement. As a comparison we also list the results of a
very simple preconditioning technique, using the identity matrix
for $\tensor{Y}_i$ in \eqref{Augustin_eq:feti_preconditioner},
where almost no reduction of the condition numbers can be noticed.

Moreover, we observe that all-floating FETI yields better
condition numbers for all preconditioners, and hence better
convergence rates of the global conjugate gradient method.
Although the global iterative method converges in less iterations
for this approach, we achieve lower computational time for the
classical FETI method for the linear elastic case with
$\mathcal{P}_1$ elements. This is mainly due to the larger
expenditure of time to set up the all-floating FETI system, the
larger coarse matrix $\tensor{G}\tensor{G}^\top$,
cf.~\eqref{Augustin_eq:projection}, and due to the higher amount
of Lagrange multipliers.
\begin{table}[t]
\caption{Iteration numbers (it.), condition numbers and
computational time (in s) for each preconditioning technique using
$\mathcal{P}_1$ elements; $\ell$ is the level of uniform
refinement.
For the L2 error the definition is given in
\eqref{eq:L2Error}, while for the estimated error of
convergence eoc the definition is given in
\eqref{eq:L2Errora}.}
\centering \small
\begin{tabular}{r|rrr|rrr|rrr|rr}
\toprule
\multicolumn{12}{l}{\textbf{all-floating}}     \\
\multicolumn{1}{c}{$\ell$}  & \multicolumn{3}{c}{identity prec.}& \multicolumn{3}{c}{lumped prec.}&\multicolumn{3}{c}{Dirichlet prec.}& L2 error & eoc \\
 1 & 61  it.\hspace*{-2mm} & 53.6\hspace{-2mm}  &  20.9 s   & 27 it.\hspace{-2mm}  &10.3\hspace{-2mm}&  19.7 s & 21 it.\hspace{-2mm} & 7.6\hspace{-2mm}&  19.5 s & 1.42E-04\hspace{-4mm}& -     \\
 2 & 71  it.\hspace*{-2mm} & 70.0\hspace{-2mm}  &  19.6 s   & 38 it.\hspace{-2mm}  &19.7\hspace{-2mm}&  18.8 s & 26 it.\hspace{-2mm} &10.4\hspace{-2mm}&  18.4 s & 3.71E-05\hspace{-4mm}& 1.94  \\
 3 & 88  it.\hspace*{-2mm} & 108.8\hspace{-2mm}  &  21.7 s   & 45 it.\hspace{-2mm}  &26.1\hspace{-2mm}&  22.3 s & 27 it.\hspace{-2mm} & 9.7\hspace{-2mm}&  22.3 s & 9.40E-06\hspace{-4mm}& 1.98  \\
 4 & 119 it.\hspace*{-2mm} & 216.8\hspace{-2mm}  &  28.8 s   & 62 it.\hspace{-2mm}  &53.2\hspace{-2mm}&  26.4 s & 32 it.\hspace{-2mm} &13.1\hspace{-2mm}&  26.6 s &  2.37E-06\hspace{-4mm}& 1.99  \\
 5 & 160 it.\hspace*{-2mm} & 432.7\hspace{-2mm}  & 116.6 s   & 91 it.\hspace{-2mm}  &126.2\hspace{-2mm}&  99.0 s & 37 it.\hspace{-2mm} &16.8\hspace{-2mm}&  105.9 s & 5.96E-07\hspace{-4mm}& 1.99  \\
\midrule
\multicolumn{12}{l}{\textbf{classical}}     \\
\multicolumn{1}{c}{$\ell$}  & \multicolumn{3}{c}{identity prec.}& \multicolumn{3}{c}{lumped prec.}&\multicolumn{3}{c}{Dirichlet prec.}& L2 error & eoc \\
1 &  80 it.\hspace{-2mm} & 98.2\hspace{-2mm} & 7.1     s& 35  it.\hspace{-2mm}  & 14.1\hspace{-2mm}& 5.9   s&29 it.\hspace{-1.5mm}  &10.0\hspace{-2mm}  &  5.9   s &  1.47E-04\hspace{-4mm}& -     \\
 2 & 105 it.\hspace{-2mm} &161.4\hspace{-2mm} & 7.8    s& 58  it.\hspace{-2mm}  & 41.9\hspace{-2mm}& 6.1   s&37 it.\hspace{-1.5mm}  &16.4\hspace{-2mm}  &  5.8   s &  3.72E-05\hspace{-4mm}& 1.98  \\
 3 & 140 it.\hspace{-2mm} &295.7\hspace{-2mm} & 9.3    s& 85  it.\hspace{-2mm}  &105.9\hspace{-2mm}& 7.9   s&46 it.\hspace{-1.5mm}  &25.4\hspace{-2mm}  &  7.7   s &  9.41E-06\hspace{-4mm}& 1.98  \\
 4 & 188 it.\hspace{-2mm} &580.9\hspace{-2mm} & 15.2   s& 125 it.\hspace{-2mm}  &252.1\hspace{-2mm}&13.1   s&54 it.\hspace{-1.5mm}  &35.8\hspace{-2mm}  &  12.2  s &  2.37E-06\hspace{-4mm}& 1.99  \\
 5 & 251 it.\hspace{-2mm} &1150.3\hspace{-2mm} & 103.4  s& 179 it.\hspace{-2mm}  &555.7\hspace{-2mm}& 88.2  s&60 it.\hspace{-1.5mm} &46.3\hspace{-2mm}  &  83.6  s &  5.96E-07\hspace{-4mm}& 1.99  \\
\bottomrule
\end{tabular}
  \label{Augustin_tab:cube512_linear}
\end{table}

From level 4, with a maximum of $8\,907$ local degrees of
freedom, to level 5, with a maximum of $66\,195$ local degrees of freedom, we
observe an increase in the local assembling and factorization time from
approximately $1.8$ seconds up to about $13$ seconds for all kinds of
preconditioners. This is mainly due to the higher memory requirements of the
direct solver. Note also that the factorization of the local stiffness matrices by the direct solver is unfeasible, if the
number of local degrees of freedom gets too large. The reason for that are memory
limitations on the VSC2 cluster. A possibility to overcome this problem is the
use of fast local iterative solvers, e.g., the CG method with a multigrid or a BPX preconditioner.
Summing it up seems that the simple lumped preconditioner and the
classical FETI approach appear to be favorable for this academic
example, with very structured subdomains and the boundary
$\Gamma_\mathrm{D}=\partial\Omega$. The latter yields a large
number of floating subdomains for all-floating FETI which are
non-floating for the classical FETI approach, and hence a much
larger coarse matrix $\tensor{G}\tensor{G}^\top$ for all-floating
FETI. The inversion of this matrix is the most time consuming part
for the levels $\ell=1,\dots,4$ that also results in the higher
computational time for all-floating FETI in these cases.

Next, we consider a linear elastic problem by using tetrahedral elements and
quadratic ansatz functions, i.e. $\mathcal{P}_2$ elements for the same mesh
and parameter properties as above. The number of degrees of freedom now varies
between $53\,181$ (level $\ell=1$) and $26\,398\,269$ (level $\ell=4$) and the number of
Lagrange multipliers between $77\,700$ and $2\,908\,692$.
Note that for all preconditioning types and for both the all-floating and the
classical FETI method the L2 error compared to the fundamental solution
behaves as predicted from the theory as we get a cubic convergence rate, see
Table~\ref{Augustin_tab:cube512_quadratic}.

For all-floating FETI we have the very interesting case that the global CG
iteration numbers remain almost constant for the lumped preconditioner, and it even
seems to be a decay for the identity and the Dirichlet preconditioner, if we increase
the local degrees of freedom, i.e. increase the refinement level $\ell$.

For the classical FETI approach the iteration numbers stay almost
constant for the Dirichlet preconditioner and increase marginally
for the other two preconditioning techniques. Concerning the
computational time we have an analogous result as in the previous
case with linear ansatz functions: the classical approach with the
lumped preconditioner seems to be the best choice for this
particular example.
\begin{table}[t]
\caption{Iteration numbers (it.), condition numbers and
computational time (in s) for each preconditioning technique using
$\mathcal{P}_2$ elements; $\ell$ is the level of uniform
refinement.
For the L2 error the definition is given in
\eqref{eq:L2Error}, while for the estimated error of
convergence eoc the definition is given in
\eqref{eq:L2Errora}. } \centering \small
\begin{tabular}{r|rrr|rrr|rrr|rr}
\toprule
\multicolumn{12}{l}{\textbf{all-floating}}     \\
\multicolumn{1}{c}{$\ell$}  & \multicolumn{3}{c}{identity prec.}& \multicolumn{3}{c}{lumped prec.}&\multicolumn{3}{c}{Dirichlet prec.}& L2 error & eoc \\
1 & 149 it. & 444.7 & 23.3 s & 73 it.\hspace{-2mm}& 73.7 & 22.0 s & 47 it. & 36.7 & 18.7 s  & 1.13E-05  & - \\
 2 & 129 it. & 330.8 & 21.9 s & 75 it.\hspace{-2mm}& 74.3 & 20.8 s & 43 it. & 27.7 & 19.3 s  & 1.44E-06  & 2.97    \\
 3 & 114 it. & 210.3 & 30.3 s & 73 it.\hspace{-2mm}& 68.8 & 27.3 s & 36 it. & 16.6 & 28.5 s  & 1.81E-07  & 2.99    \\
 4 & 105 it. & 167.8 & 99.8 s & 69 it.\hspace{-2mm}& 65.2 & 93.4 s & 33 it. & 14.4 & 90.2 s  & 2.26E-08  & 3.00   \\
\midrule
\multicolumn{12}{l}{\textbf{classical}}     \\
\multicolumn{1}{c}{$\ell$}  & \multicolumn{3}{c}{identity prec.}& \multicolumn{3}{c}{lumped prec.}&\multicolumn{3}{c}{Dirichlet prec.}& L2 error & eoc \\
1 & 120 it. & 405.0 &  7.5 s &  65 it.\hspace{-2mm} & 48.9 &  6.9 s & 40 it. & 21.0 &  6.5 s  & 1.17E-05    & -\\
 2 & 108 it. & 302.6 &  7.5 s &  69 it.\hspace{-2mm} & 57.6 &  6.7 s & 41 it. & 20.6 &  7.5 s  & 1.46E-06    & 3.00  \\
 3 & 112 it. & 253.4 & 12.6 s &  91 it.\hspace{-2mm} & 116.2& 11.7 s & 42 it. & 21.0 & 12.3 s  & 1.82E-07    & 3.01  \\
 4 & 136 it. & 273.1 & 76.3 s & 128 it.\hspace{-2mm} & 262.8& 77.3 s & 48 it. & 27.7 & 79.1 s  & 2.26E-08    & 3.01   \\
\bottomrule
\end{tabular}
  \label{Augustin_tab:cube512_quadratic}
\end{table}
\begin{figure}[htbp]
\centering\tiny
\includegraphics[width=0.43\textwidth]{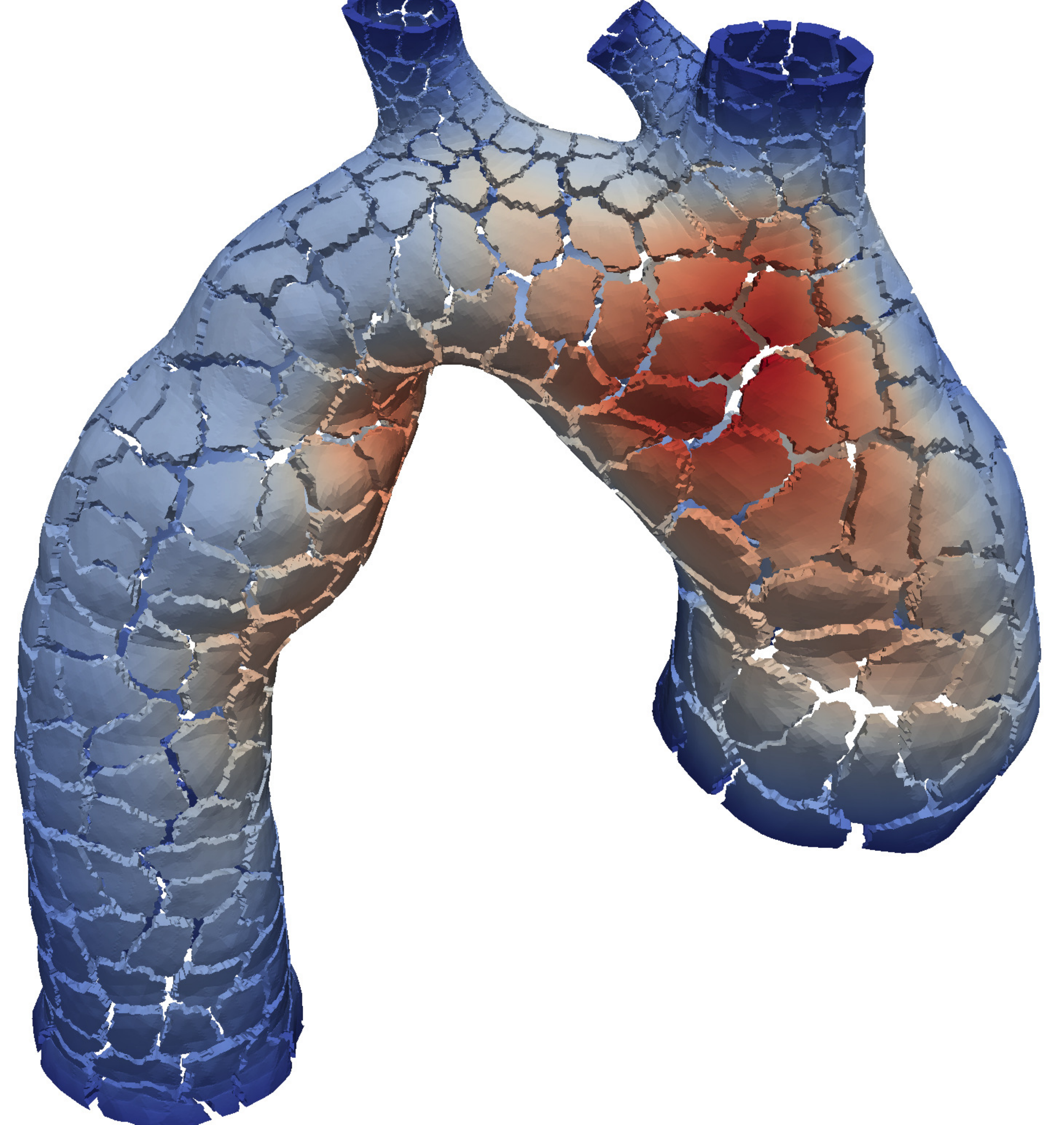}
\hspace{-10pt}
\def\svgwidth{74mm}\input{augustin_aorta_points.tex}
\caption{ Mesh of an aorta seen from above showing
  the brachiocephalic artery, and the left common carotid and subclavian arteries.
  The fine mesh consists of $5\,418\,594$ tetrahedrons and $1\,055\,901$ vertices, while colors indicate the displacement field
  with an internal pressure of $1$\,mmHg.
  Additionally, the splits show the decomposition of the mesh into $480$ subdomains (left).
  Coarser mesh consisting of $720\,060$ tetrahedrons and $150\,725$ vertices used in Section~\ref{sec:loadStepping} with $5$ selected
  vertices A--E (right); colors show the distribution of the stress magnitude $\tensor{\sigma}_\mathrm{mag}$ according to (\ref{stressmag})
  with an internal pressure of $300$\,mmHg.
  For both images red indicates high and blue low values.} \label{Augustin_fig:aorta}
\centering
    \includegraphics[width=0.47\textwidth] {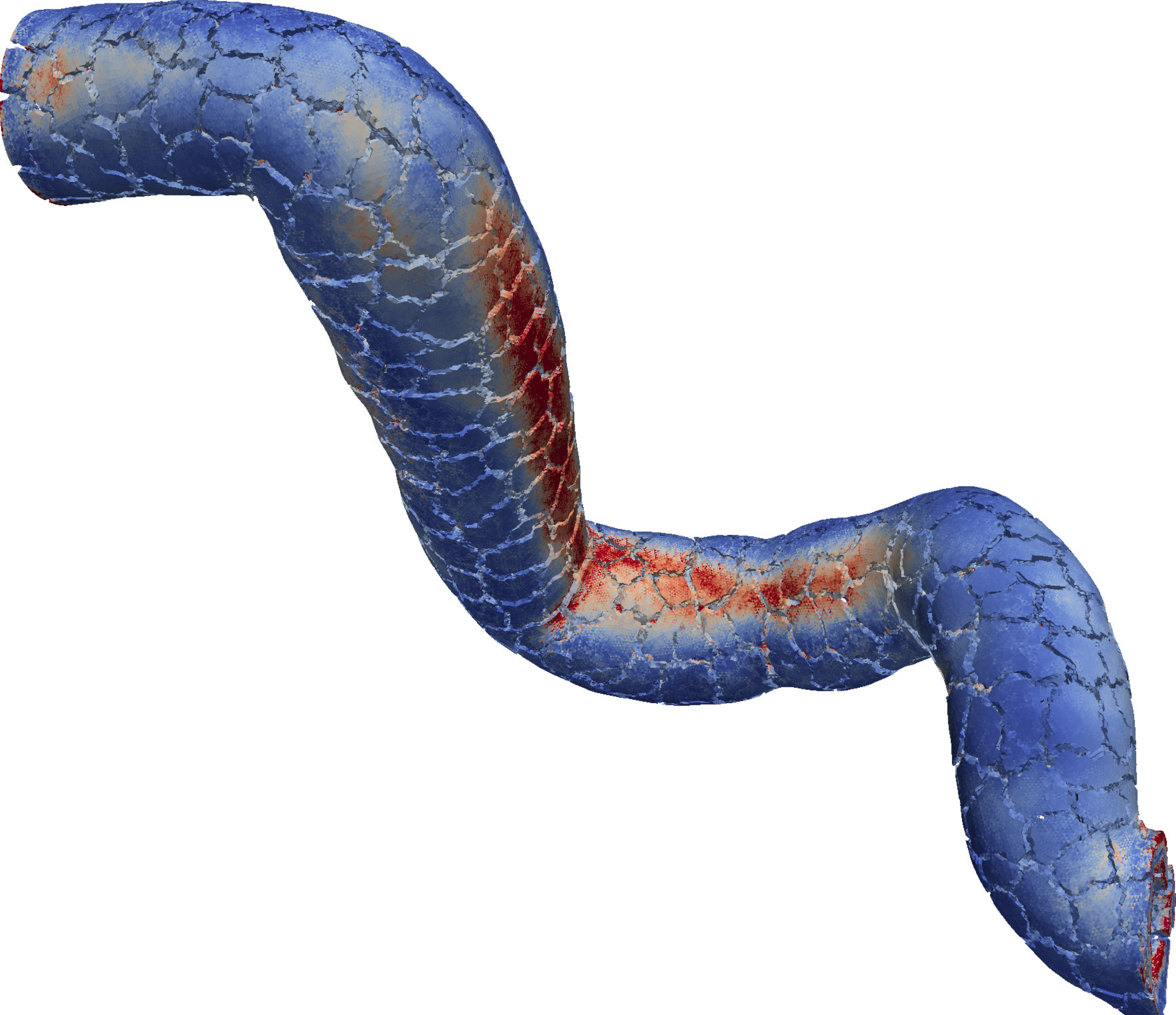}\hspace{5pt}
    \includegraphics[width=0.38\textwidth]{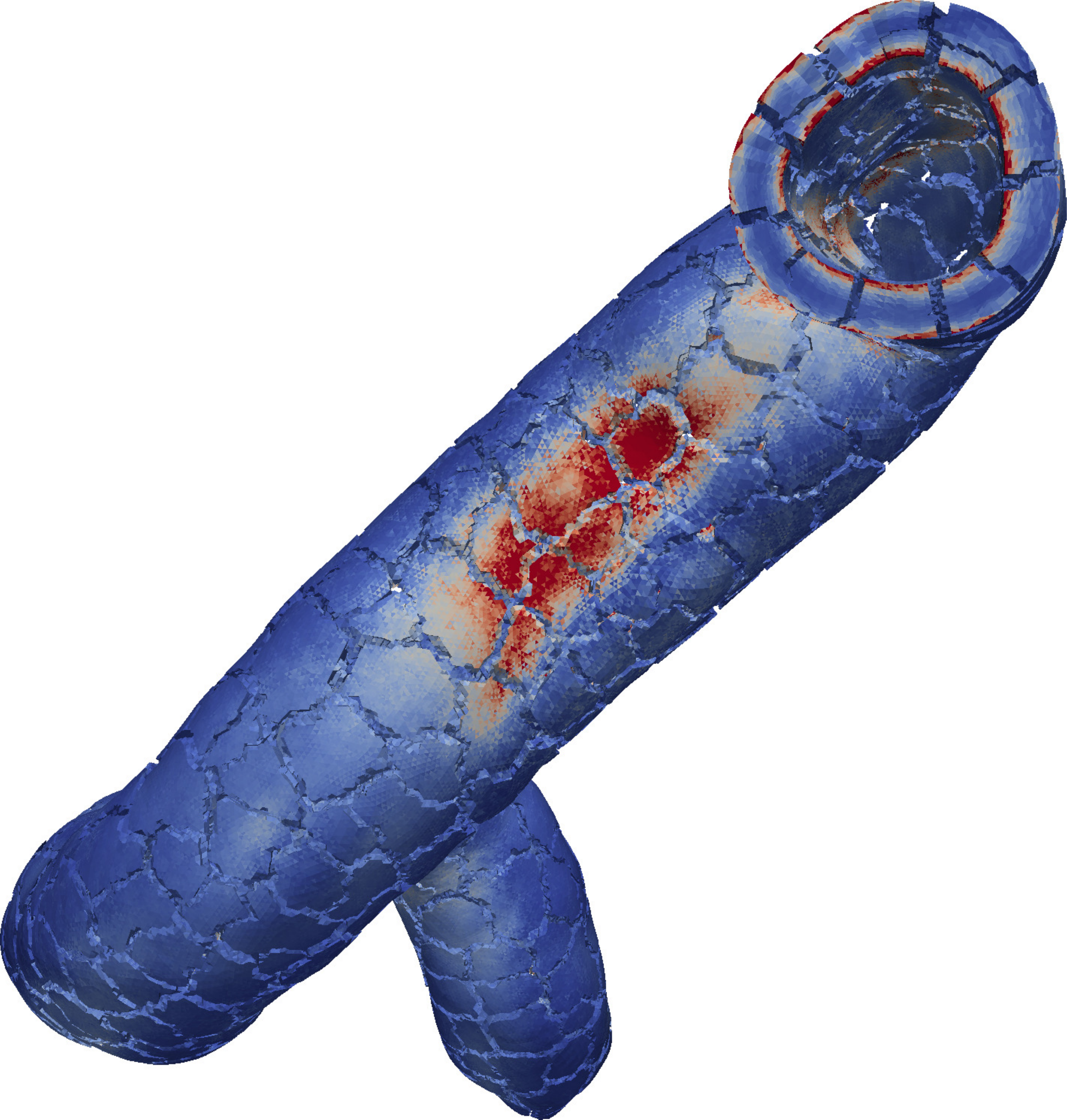}
\caption{Mesh of a segment of a common carotid artery from two
  different points of view. The mesh consists of $9\,195\,336 $ tetrahedrons and $1\,621\,365$ vertices.
  Color indicates the distribution of the stress magnitude $\tensor{\sigma}_\mathrm{mag}$ according to (\ref{stressmag})
  due to an internal pressure of $1$\,mmHg, red indicates high and blue low values.
  Additionally, the splits show the decomposition of the mesh into $512$ subdomains.}
  \label{Augustin_fig:carotis}
\end{figure}
\begin{figure}[htbp]
  \centering
    \includegraphics[width=0.45\textwidth]{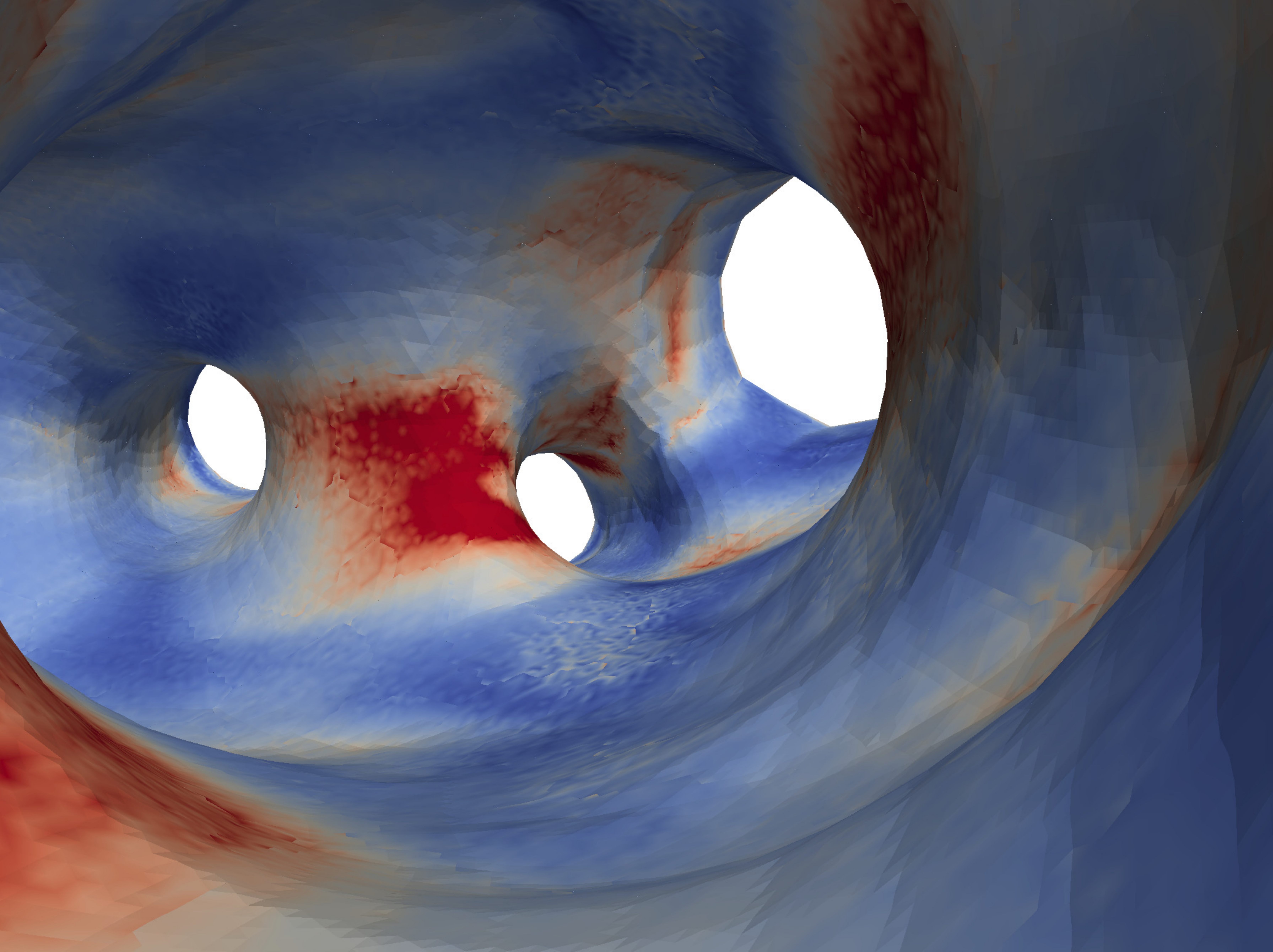}\hspace{0.5cm}
    \includegraphics[width=0.45\textwidth]{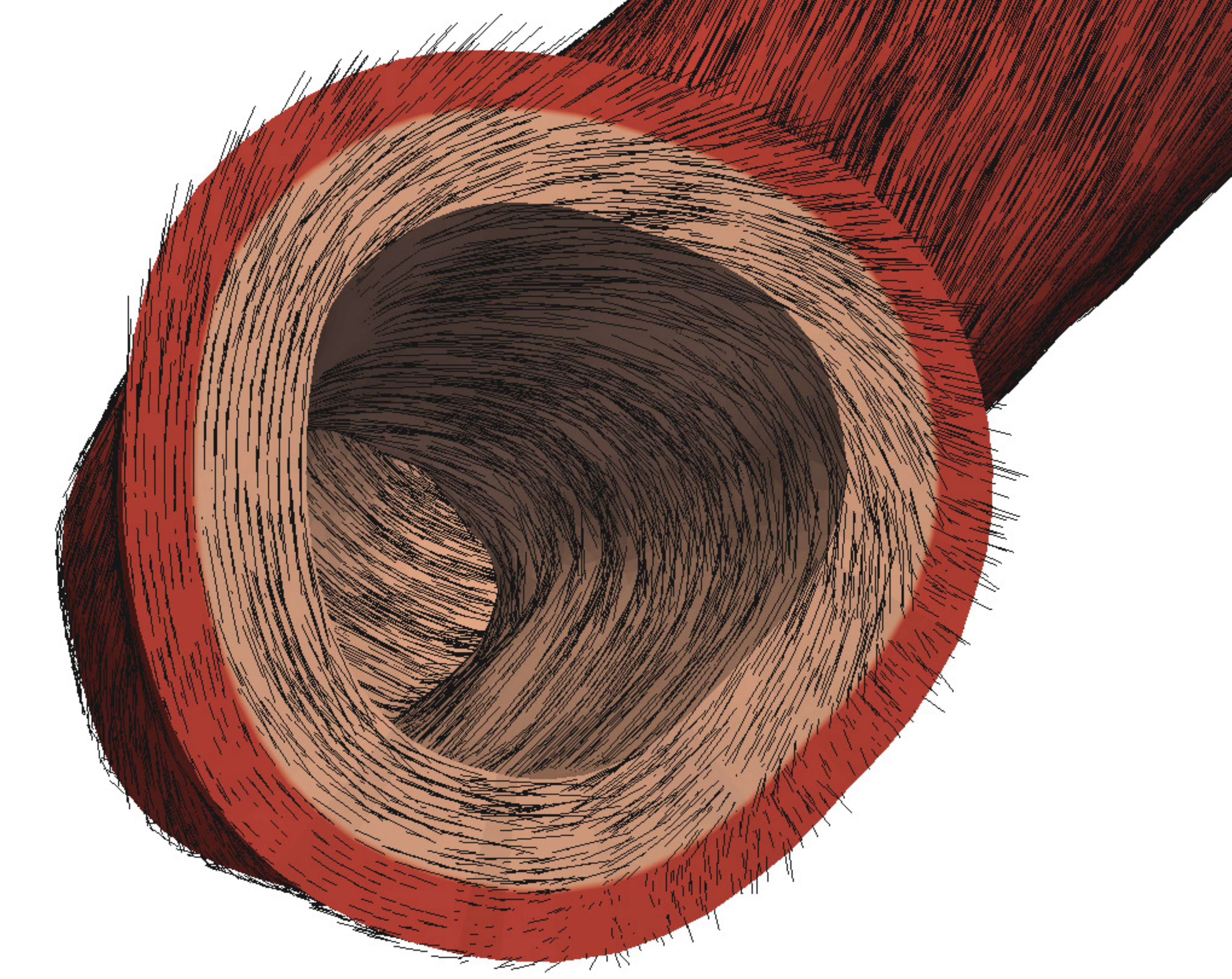}
  \caption{Distribution of the stress magnitude $\tensor{\sigma}_\mathrm{mag}$ inside the aorta (left);
    values of high stress in red and of low stress in blue. To the right the fiber directions
    (black curves) and the two layers (adventitia in red and media in orange) of the
    carotid artery are shown.}
    \label{Augustin_fig:fiberDirections}
  \end{figure}
\subsection{Arterial model on a realistic mesh geometry}\label{Augustin_sec:arterialExamples}
In this section we present examples to show the applicability of
the FETI approaches for biomechanical applications, in particular
the inflation of an artery segment. We consider the mesh of an
aorta and the mesh of a common carotid artery, see
Figs.~\ref{Augustin_fig:aorta} and \ref{Augustin_fig:carotis}. The
geometries are from AneuriskWeb \cite{Augustin_aneurisk} and Gmsh
\cite{Augustin_geuzaine2009}. The generation
of the volume mesh was performed using VMTK
and Gmsh \cite{Augustin_geuzaine2009}.

The fiber directions, see Fig.~\ref{Augustin_fig:fiberDirections}
(right), were calculated using a method described by Bayer et al.
\cite{Augustin_bayer2012} for the myocardium. To adapt this method
for the artery we first solved the Laplace equation on the domain
$\Omega_0$ with homogeneous Dirichlet boundary conditions on the
inner surface and inhomogeneous Dirichlet boundary conditions on
the outer wall. The gradient of the solution is used to define the
transmural direction $\hat{\vec{e}}_2$ in each element. As a
second step we repeat this procedure using homogeneous Dirichlet
boundary conditions on the inlet surface and inhomogeneous
boundary conditions on the outlet surfaces which yields the
longitudinal direction $\hat{\vec{e}}_1$. The cross product of
these two vectors eventually provides the circumferential
direction $\hat{\vec{e}}_0$. With a rotation we get the two
desired fiber directions $\vec{a}_{0,1}$ and $\vec{a}_{0,2}$  in
the media and the adventitia, respectively. Thus,
\begin{equation}\label{Augustin_eq:rotation}
  \begin{pmatrix}
    \vec{a}_{0,1} & -\vec{a}_{0,2} & \hat{\vec{e}}_2\end{pmatrix}=
  \begin{pmatrix}
    \hat{\vec{e}}_0&\hat{\vec{e}}_1& \hat{\vec{e}}_2\end{pmatrix}
  \begin{pmatrix}
    \cos \alpha & -\sin\alpha & 0\\
    \sin \alpha & \cos\alpha  & 0\\
    0&0&1
  \end{pmatrix}
  \begin{pmatrix}
    \hat{\vec{e}}_0^\top\\\hat{\vec{e}}_1^\top\\ \hat{\vec{e}}_2^\top\end{pmatrix}
  \begin{pmatrix}
    \hat{\vec{e}}_0&\hat{\vec{e}}_1& \hat{\vec{e}}_2\end{pmatrix}.
\end{equation}
The value for the angle $\alpha$ are $29^\circ$ for the media and
$62^\circ$ for the adventitia, taken from
\cite{Augustin_holzapfel2000b}.

To describe the anisotropic and nonlinear arterial
tissue, we use the material model
(\ref{Augustin_eq:isoaniso}--\ref{Augustin_eq:aniso_2}), with
the parameters given in Table~\ref{Augustin_tab:param} and
$\kappa$ is varied. Dirichlet boundary conditions
\eqref{Augustin_eq:Dirichlet} are imposed on the respective
intersection areas. We perform an inflation simulation on the
artery segment where the interior wall is exposed to a constant
pressure $p$. This is performed using Neumann boundary conditions
\eqref{Augustin_eq:Neumann}. If not stated otherwise, we present
the results of one load step applying a rather low pressure of
$1$\,mmHg. This is necessary to have a converging Newton method.
Nonetheless, the material model as used is anisotropic. To
simulate a higher pressure, an appropriate load stepping scheme,
see \eqref{Augustin_eq:loadStepping},
 has to be used. However, this does not affect
the number of local iterations significantly.  As already
mentioned in Section~\ref{Augustin_sec:FETI} we use the CG method
as global iterative solver. Experiments with a standard
non-symmetric nonlinear elasticity system and the necessary GMRES
method as an iterative solver showed similar results, as presented
in the following with the symmetric system. However, the memory
requirements of the GMRES solver are much higher.

The local generalized pseudo-inverse matrices are realized with a
sparsity preserving regularization by fixing nodes,
see, e.g., \cite{Augustin_brzobohaty2011}, and the direct solver
package Pardiso. The global nonlinear finite element system is
solved by a Newton scheme, where the FETI approach is used in each
Newton step. For the considered examples the Newton scheme needed
four to six iterations. Due to the non-uniformity of the
subdomains the efficiency of a global preconditioner becomes more
important. It may happen that the decomposition of a mesh results
in subdomains that have only a few points on the Dirichlet
boundary. This negatively affects the convergence of the CG method
using classical FETI, but does not affect the global iterative
method of the all-floating approach at all. This is a major
advantage of all-floating FETI since here all subdomains are
treated the same, and hence all subdomains are stabilized. This
behavior is observed for almost all settings for preconditioners
and the penalty parameter $\kappa$ as well as for linear and
quadratic ansatz functions, see
Tables~\ref{Augustin_tab:aorta_linear}--\ref{Augustin_tab:carotis_quadratic}.
\begin{table}[htpb]
\caption{Iteration numbers (it.) per Newton step and computational
time (in s) per Newton step for the all-floating and the
classical FETI approach with \textit{linear} ansatz functions
comparing the three considered preconditioners. The penalty
parameter $\kappa$ was varied from $10$ to $1000$\,kPa.
Mesh: mesh of the aorta subdivided in $480$ subdomains, computed
with $480$ cores.}
  \centering
  \small
\begin{tabular}{r|rrr|rrr|rcl}
\toprule
\multicolumn{10}{l}{\textbf{all-floating}}     \\
 $\kappa$  & \multicolumn{3}{c}{identity preconditioner}&  \multicolumn{3}{c}{lumped preconditioner}
 &\multicolumn{3}{c}{Dirichlet preconditioner} \\
10   & 1052 it. & & 57.6 s & 160 it.& & 31.0 s & 56 it. & & 22.8 s      \\
100  & 1879 it. & & 94.6 s & 305 it.& & 29.5 s & 85 it. & & 25.4 s      \\
1000 & 4122 it. & &177.1 s & 681 it.& & 48.8 s &209 it. & & 31.8 s      \\
\midrule
\multicolumn{10}{l}{\textbf{classical}}     \\
 $\kappa$  & \multicolumn{3}{c}{identity preconditioner}&  \multicolumn{3}{c}{lumped preconditioner}
 &\multicolumn{3}{c}{Dirichlet preconditioner} \\
10   & 2056 it. & & 98.7 s & 305 it.& & 35.5 s &117 it. & & 27.2 s      \\
100  & 3711 it. & &149.8 s & 540 it.& & 35.5 s &144 it. & & 28.4 s      \\
1000 & 8245 it. & &327.8 s &1190 it.& & 60.9 s &263 it. & & 32.9 s      \\
\bottomrule
\end{tabular}
  \label{Augustin_tab:aorta_linear}
\end{table}
\begin{table}[htpb]
\caption{Iteration numbers (it.) per Newton step and computational
time (in s) per Newton step for the all-floating and the
classical FETI approach with \textit{linear} ansatz functions
comparing the three considered preconditioners. The penalty
parameter $\kappa$ was set to $1000$\,kPa. Mesh: mesh of the
carotid artery with two layers (adventitia and
media) subdivided in $512$ subdomains, computed with $512$ cores.}
  \centering
\small
\begin{tabular}{r|rrr|rrr|rcl}
\toprule
 type  & \multicolumn{3}{c}{identity preconditioner}&  \multicolumn{3}{c}{lumped preconditioner}
 &\multicolumn{3}{c}{Dirichlet preconditioner} \\
all-floating & $>$ 10000 it. & & - s   & 1084 it.& & 100.6 s & 497 it. & & 85.5 s      \\
classical    & 5130 it.      & & 357 s & 1794 it.& & 200.2 s & 588 it. & & 97.7 s      \\
\bottomrule
\end{tabular}
  \label{Augustin_tab:carotis_linear}
\end{table}
\begin{table}[htpb]
\caption{Iteration numbers (it.) per Newton step and computational
time (in s) per Newton step for the all-floating and the
classical FETI approach with \textit{quadratic} ansatz functions
comparing the three considered preconditioners. The penalty
parameter $\kappa$ was varied from $10$ to $1000$\,kPa.
Mesh: mesh of the aorta subdivided in $480$ subdomains, computed
with $480$ cores.}
 \centering \small
\begin{tabular}{r|rrr|rrr|rcl}
\toprule
\multicolumn{10}{l}{\textbf{all-floating}}     \\
 $\kappa$  & \multicolumn{3}{c}{identity preconditioner}&  \multicolumn{3}{c}{lumped preconditioner}
 &\multicolumn{3}{c}{Dirichlet preconditioner} \\
10   &  940 it. & &  491.1 s & 283 it.& &209.5 s & 71 it. & &157.3 s      \\
100  & 1519 it. & & 1186.4 s & 523 it.& &332.0 s &105 it. & &178.1 s      \\
1000 & 3371 it. & & 2584.5 s &1372 it.& &746.0 s &206 it. & &282.7 s      \\
\midrule
\multicolumn{10}{l}{\textbf{classical}}     \\
 $\kappa$  & \multicolumn{3}{c}{identity preconditioner}&  \multicolumn{3}{c}{lumped preconditioner}
 &\multicolumn{3}{c}{Dirichlet preconditioner} \\
10   & 1319 it. & & 654.2 s & 333 it.& &225.2 s &113 it. & &188.4 s      \\
100  & 2362 it. & &1140.6 s & 664 it.& &402.6 s &110 it. & &177.5 s      \\
1000 & 5563 it. & &4168.3 s &1742 it.& &943.1 s &204 it. & &280.1 s      \\
\bottomrule
\end{tabular}
  \label{Augustin_tab:aorta_quadratic}
\end{table}
\begin{table}[htpb]
\caption{Iteration numbers (it.) per Newton step and computational
time (in s) per Newton step for the all-floating and the
classical FETI approach with \textit{quadratic} ansatz functions
comparing the three considered preconditioners. The penalty
parameter $\kappa$ was set to $1000$\,kPa. Mesh: mesh of the
carotid artery with two layers (adventitia and
media) subdivided in $1024$ subdomains, calculated with $1024$
cores. }
  \centering
\small
\begin{tabular}{r|rrr|rrr|rcl}
\toprule
 type  & \multicolumn{3}{c}{identity preconditioner}&  \multicolumn{3}{c}{lumped preconditioner}
 &\multicolumn{3}{c}{Dirichlet preconditioner} \\
all-floating & $>$ 10000 it. & & - s     & 2163 it.& &1133.9 s & 674 it. & &994.6 s      \\
classical    & 6006 it.      & &2672.6 s & 4798 it.& &2306.8 s & 764 it. & &771.2 s      \\
\bottomrule
\end{tabular}
  \label{Augustin_tab:carotis_quadratic}
\end{table}

For example, applying all-floating FETI with the Dirichlet
preconditioner to the mesh of the aorta using a penalty parameter
$\kappa=1000$\,kPa the global CG method converged in considerable
less iterations (209) than the CG method using classical FETI
(263), see Table~\ref{Augustin_tab:aorta_linear}. The advantage of
the smaller number of iterations is not so significantly reflected
in the computational time since, as for the linear case, we have
higher set up times and a larger coarse system
$\tensor{G}\tensor{G}^\top$. Nonetheless, for the considered
examples it shows that all-floating FETI yields
lower iteration numbers of the global systems and it
is also competitive or even advantageous with
respect to the classical approach concerning the computational
time.

In contrast to the academic example in
Section~\ref{Augustin_sec:NRLinElast} the more complex Dirichlet
preconditioner is the best choice for all considered settings.
Especially for $\kappa\gg 1$ the iteration numbers with the lumped
and the identity preconditioner escalate. Admittedly, the numbers
in Table~\ref{Augustin_tab:aorta_linear} also show that the
convergence of the CG method, within all FETI approaches and
preconditioner settings, is dependent on the penalty parameter
$\kappa$.

Using quadratic ansatz functions we have a total number of
$23\,031\,620$ degrees of freedom for the aorta mesh and
$36\,527\,435$ degrees of freedom for the carotid
artery mesh. In order to not infringe the memory limitations on
the \textit{VSC2} cluster we have to use a decomposition into
$1024$ subdomains (instead of $512$) for the carotid
artery. For the aorta it was possible to stay with $480$
subdomains. The number of Lagrange multipliers are then
$1\,552\,665$ (aorta) and $4\,585\,203$ (carotid
artery). Comparing the numbers in
Table~\ref{Augustin_tab:aorta_quadratic} and
Table~\ref{Augustin_tab:carotis_quadratic} show similar results as
in the case with linear ansatz functions. The Dirichlet
preconditioner is preferable for all test cases and the
all-floating approach is competitive to the classical FETI
approach. Albeit quadratic ansatz functions resolve the nearly
incompressible elastic behavior better than linear ansatz
functions we also notice a correlation between the global
iteration numbers and the penalty parameter $\kappa$, see
Table~\ref{Augustin_tab:aorta_quadratic}. Nonetheless, the
iteration numbers do not increase as much as for the
$P_1-P_0$ element case and the values of $J=\det\tensor{F}$ in
each element are much closer to $1$ for the
$P_2-P_0$ elements.
\subsection{Load stepping scheme}\label{sec:loadStepping}
In this section we analyze the biomechanical behavior of the aorta
up to an internal pressure of $300$\,mmHg. Higher pressures would
induce damage and softening behavior which cannot be captured with
the arterial model discussed in Section~\ref{Augustin_sec:model}.
For that purpose we consider a coarser version of the mesh of the
aorta (see Fig.~\ref{Augustin_fig:aorta}), which is subdivided
into $32$ subdomains since for this mesh the all-floating FETI
method looks significantly advantageous.
The reasons for that are as follows: (i) we have lower iteration
numbers for the all-floating FETI approach, as already observed
in Section~\ref{Augustin_sec:arterialExamples}; (ii) the matrix
$\tensor{G}\tensor{G}^\top$ in \eqref{Augustin_eq:projection} is
small, and hence less time is needed to compute the inverse of
this coarse system, especially in comparison to the assembly time
and the global solving time of the CG method.

With this mesh we simulate an arterial model with the parameters
from Table~\ref{Augustin_tab:param} and with $c=6$\,kPa and
$\kappa=1000$\,kPa using the Dirichlet preconditioner. The results
of a load stepping scheme, where we applied an internal pressure
up to $300$\,mmHg over $572$ loading steps, are found in the
Figs.~\ref{Augustin_fig:timestepping} and
\ref{Augustin_fig:timesteppingIterations}. Note that the average
iteration number over one time step increased from $248$ to $268$
for all-floating FETI and from $340$ to $358$ for the classical
FETI approach for higher pressures, and, consequently, a more
anisotropic material behavior. The simulation needed four to five
Newton steps and the solving times for all-floating FETI are
significantly faster, see
Fig.~\ref{Augustin_fig:timesteppingIterations}.

In our plots we used a \emph{stress magnitude}
$\tensor{\sigma}_\mathrm{mag}$ according to
\begin{equation}
  \tensor{\sigma}_\mathrm{mag} =
  \sqrt{\sigma_{11}^2+\sigma_{22}^2+\sigma_{33}^2+2\sigma_{12}^2+2\sigma_{13}^2+2\sigma_{23}^2},
  \label{stressmag}
\end{equation}
used as a measure to visualize our data. For advantages and
disadvantages of certain stress values concerning the analysis of
rupture and failure in aortic tissues, see, e.g.,
\cite{Augustin_humphrey2012}. Other values used in
Fig.~\ref{Augustin_fig:timestepping} are the \emph{displacement
norm} $u_{\mathrm{norm}}$ and the \emph{relative displacement}
$u_\mathrm{rel}$, i.e.
\begin{equation}
  u_{\mathrm{norm}} = \sqrt{u_{1}^2 + u_{2}^2 + u_{3}^2},
  \qquad
  u_\mathrm{rel} = \frac{u_\mathrm{norm}}{u_\mathrm{max}},
\end{equation}
for a point with the displacement vector $\vec{u} =
(u_{1},u_{2},u_{3})$ at the time step $t$, and $u_\mathrm{max}$ is
the largest occurring displacement norm for that point over all
time steps.
\begin{figure}[!htbp]
\centering
    \hspace*{-7mm} \input{augustin_stress_vs_displacement.tex}
    \hspace*{-11mm}\input{augustin_displacement_vs_pressure.tex}\hspace{-4mm}
    \caption{Stress magnitude $\tensor{\sigma}_\mathrm{mag}$ versus
    relative displacement $u_\mathrm{rel}$ (left) and evolution of the displacement norm
    $u_{\mathrm{norm}}$ over the load steps up to an internal pressure $p$ of $300$\,mmHg (right).
    The plots were generated using data at the specific points A--E, as shown in
    Fig.~\ref{Augustin_fig:aorta} (right).}
\label{Augustin_fig:timestepping}
\vspace*{0.4cm}
\centering
    \hspace*{-7mm} \input{augustin_timestepping_iterations.tex}
    \hspace*{-11mm}\input{augustin_timestepping_solve_time.tex}\hspace{-4mm}
    \caption{Comparison of all-floating FETI (gray) and classical FETI (black) for a time stepping scheme. Average iteration numbers of one time step (left) and solving times in seconds for one time step (right) over 572 load steps.}
\label{Augustin_fig:timesteppingIterations}
\end{figure}

\subsection{Strong scaling for nonlinear elasticity}
\label{strongscaling}
Here we analyze our computational framework with
respect to strong scaling efficiency, i.e.
\begin{equation}\label{Augustin_eq:efficiency}
  \mathrm{eff}=\frac{t_\mathrm{I}}{P \,t_\mathrm{P} },
\end{equation}
where $t_\mathrm{I}$ is the amount of time to complete a computation with the initial number of
processing units $\mathrm{I}$ (in our case $\mathrm{I}=16$) and $t_\mathrm{P}$ is the amount of time to complete
the same computation with $P$ processing units.
In particular, we consider the meshes of the carotid artery and
the aorta as in Section~\ref{Augustin_sec:arterialExamples}, both
subdivided into $512$ subdomains. We apply the arterial model with
the parameters from Table~\ref{Augustin_tab:param} and use a
$\kappa = 100$ with the lumped preconditioner and linear ansatz
functions. For the aorta we used all-floating FETI and needed an
average of $324$ global CG iterations to reach an absolute error
of $\varepsilon = 10^{-8}$ and $5$ Newton steps to reach an
absolute error of $10^{-6}$. In the case of the
carotid artery and classical FETI we needed $674$ global CG
iterations and also $5$ Newton steps to reach the same error
limits as above.
\begin{table}[!htbp]
  \caption{Computational time (in s) and efficiency ($\mathrm{eff}$)
  according to \eqref{Augustin_eq:efficiency} for a
nonlinear elastic problem using a varying number of processing
units $P$. The time is measured for $1$ time step with $5$ Newton
steps for all-floating FETI and the lumped preconditioner. }
\centering \small
\begin{tabular}{*{7}{r}}
\toprule
 $P$  & local time & $\mathrm{eff}$ & global CG time & $\mathrm{eff}$ & total time & $\mathrm{eff}$ \\
\midrule
 16 &407.7  s & 1.000  &1311.7 s & 1.000     & 2028.6 s  & 1.000         \\
 32 &203.1  s & 1.004  & 666.4 s &  0.984    & 1054.2 s  & 0.962     \\
 64 &101.7  s & 1.002  & 345.4 s &  0.949    &  562.0 s  & 0.902     \\
128 & 50.5  s & 1.009  & 184.7 s &  0.888    &  316.7 s  & 0.801     \\
256 & 25.3  s & 1.007  & 103.8 s &  0.790    &  192.8 s  & 0.658     \\
512 & 12.7  s & 1.000  & 67.6  s &  0.606    &  161.0 s  & 0.394     \\
\bottomrule
\end{tabular}
  \label{Augustin_tab:aorta_scaling}
\end{table}
\begin{table}[!htbp]
\caption{Computational time (in s) and efficiency ($\mathrm{eff}$)
according to \eqref{Augustin_eq:efficiency} for a
nonlinear elastic problem on the carotid artery mesh
using a varying number of processing units $P$. The time is
measured for $1$ time steps with $5$ Newton steps for classical
FETI and the lumped preconditioner.} \centering \small
\begin{tabular}{*{7}{r}}
\toprule
 $P$  & local time & $\mathrm{eff}$ & global CG time & $\mathrm{eff}$ & total time & $\mathrm{eff}$ \\
\midrule
16  &  726.0 s   & 1.000    &  4725.8 s & 1.000   & 6519.7 s  & 1.000          \\
 32 &  351.3 s   & 1.033    &  2368.2 s & 0.998   & 3497.0 s  & 0.932      \\
 64 &  170.5 s   & 1.065    &  1262.9 s & 0.936   & 1991.2 s  & 0.819      \\
128 &   90.7 s   & 1.001    &   694.5 s & 0.851   & 1194.1 s  & 0.682        \\
256 &   47.3 s   & 0.960    &   443.6 s & 0.666   &  914.4 s  & 0.446        \\
512 &   23.9 s   & 0.949    &   297.2 s & 0.497   &  667.4 s  & 0.305        \\
\bottomrule
\end{tabular}
  \label{Augustin_tab:carotis_scaling}
\end{table}

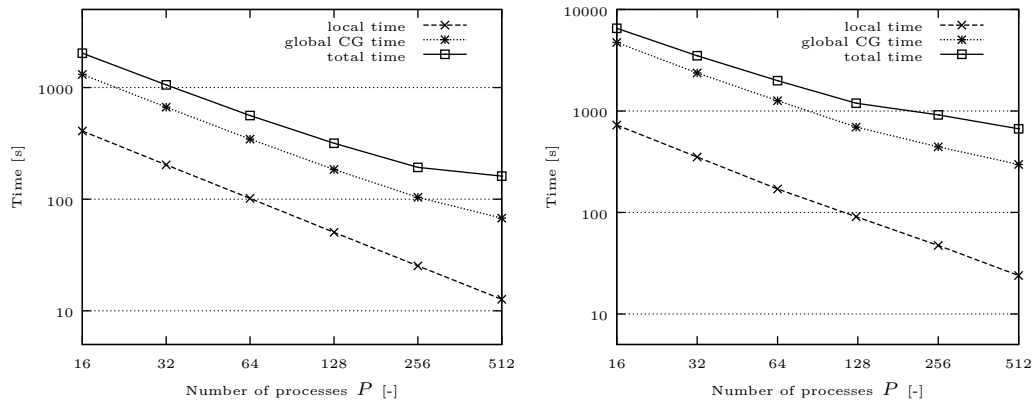
\begin{figure}[htbp]
\centering
 \hspace*{-18mm}\input{augustin_solve_time_aorta.tex}
 \hspace*{-15mm}\input{augustin_solve_time_carotis.tex}\hspace{-8mm}
 \caption{Computation times (in s) for a simulation of the
 anisotropic arterial model with the aorta mesh (left) and the carotid artery
 mesh (right) using a varying number of cores.}
\label{Augustin_fig:scaling}
\end{figure}

In the Tables~\ref{Augustin_tab:aorta_scaling} and
\ref{Augustin_tab:carotis_scaling} we present the following
numbers: the \textit{local time} is the sum of all assembling and
local factorization times during the solution steps. The
factorization of the local problems was performed with the direct
solver package Pardiso. In most cases we observed a super-linear
speedup, and hence an efficiency greater than $1$ for this value.
This is due to memory issues, mainly so-called cache effects.
For more information on this well-known phenomenon,
see, e.g., \cite{Augustin_hennessy2012}. The \textit{global CG
time} is the duration of all CG solution steps together. We see
that this value scales very well up to $256$ cores for the aorta
and up to $128$ cores for the carotid artery.
The \textit{total time} is the total computational time including
input and output functions. It also scales admissibly well up to
$256$ processing units for the aorta, and up to $128$ cores for
the carotid artery, see
Tables~\ref{Augustin_tab:aorta_scaling} and
\ref{Augustin_tab:carotis_scaling}, and
Fig.~\ref{Augustin_fig:scaling}. For a higher number of cores, at
least for the specific examples, the speedup is rather low.
Possibilities to overcome this problem are, for example, the usage
of parallel solver packages such as \textit{hypre} and a more
efficient assembling of the coarse system of the FETI method. It
also needs a more elaborate strategy with MPI and the memory
management. Note that at some point the subdomains get too small
and the increasingly dominant MPI communication impedes further
strong scaling.

\section{Discussion and Limitations}
We have shown the application of the finite element tearing and
interconnecting method to elasticity problems, in particular to
the simulation of the nonlinear elastic behavior of cardiovascular
tissues such as the artery. The main ideas of domain decomposition
methods were summarized and the classical and the all-floating
FETI approach were discussed in detail.

Illustrated by representative numerical examples we
have shown certain advantages of the all-floating FETI method
compared to the classical FETI approach. To the best of our
knowledge the application of the all-floating approach to
nonlinear anisotropic elasticity problems cannot be found in the
literature. Certainly, the mentioned advantages are
influenced by the mesh structure and the choice of the boundary
conditions, and hence the method to choose depends on the specific
problem.

We have presented and compared different techniques of preconditioning: the
lumped preconditioner and the optimal Dirichlet preconditioner.
Furthermore, the numerical examples exposed some instabilities of the global
iterative method for nearly incompressible material parameters, i.e. for a very large
penalty parameter $\kappa$.
Here we were able to present, like it was also shown in
earlier contributions, that quadratic ansatz functions resolve the incompressible
 elastic behavior better than linear ansatz functions.

\section*{Acknowledgements}
This work was supported by the Austrian Science Fund (FWF)
and by Graz University of Technology within the SFB
Mathematical Optimization and Applications in Biomedical Sciences.
The authors would like to thank Dr.~G\"unther~Of, Graz University
of Technology and Dr.~Clemens~Pechstein, Johannes Kepler
University of Linz, for the fruitful cooperation and many helpful
discussions.
\clearpage

\end{document}

%% file: augustin_artery.tex
\begingroup%
  \makeatletter%
  \providecommand\color[2][]{%
    \errmessage{(Inkscape) Color is used for the text in Inkscape, but the package 'color.sty' is not loaded}%
    \renewcommand\color[2][]{}%
  }%
  \providecommand\transparent[1]{%
    \errmessage{(Inkscape) Transparency is used (non-zero) for the text in Inkscape, but the package 'transparent.sty' is not loaded}%
    \renewcommand\transparent[1]{}%
  }%
  \providecommand\rotatebox[2]{#2}%
  \ifx\svgwidth\undefined%
    \setlength{\unitlength}{348.696417bp}%
    \ifx\svgscale\undefined%
      \relax%
    \else%
      \setlength{\unitlength}{\unitlength * \real{\svgscale}}%
    \fi%
  \else%
    \setlength{\unitlength}{\svgwidth}%
  \fi%
  \global\let\svgwidth\undefined%
  \global\let\svgscale\undefined%
  \makeatother%
  \begin{picture}(0.5,1.54663042)%
    \put(0,0.05){\includegraphics[width=\unitlength]{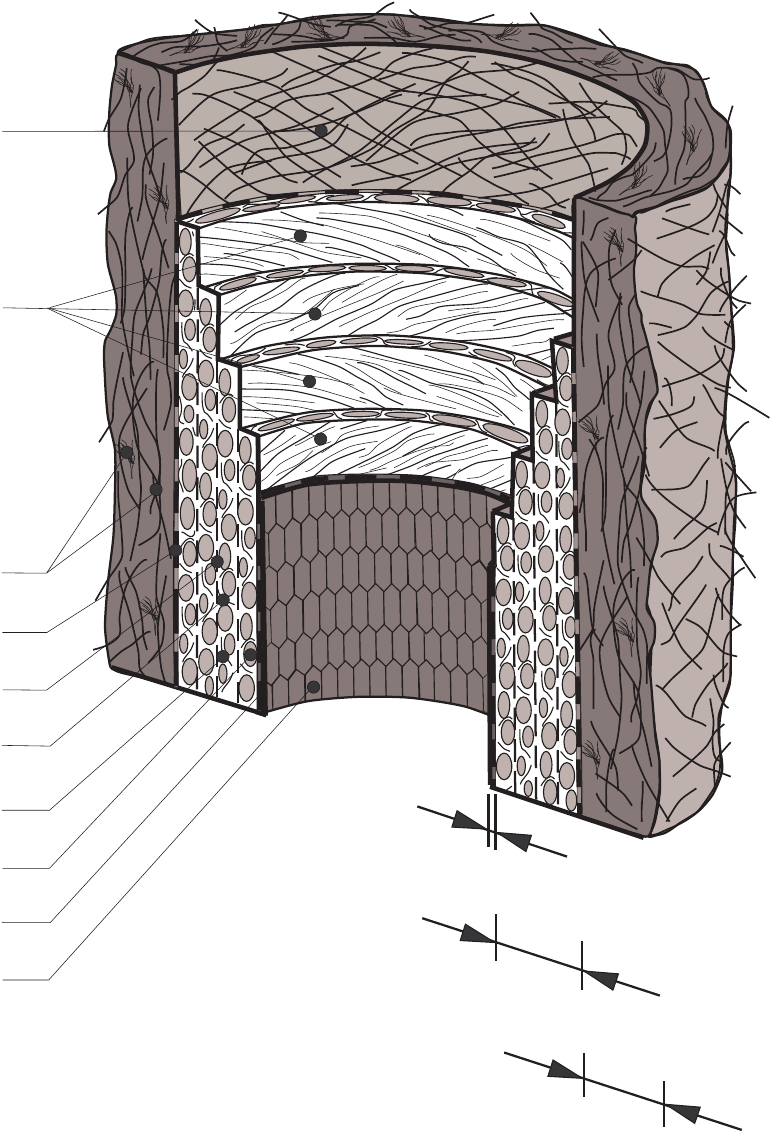}}%
    \put(0.29853022,0.77734481){\makebox(0,0)[lb]{\smash{1}}}%
    \put(0.58117867,0.39253361){\rotatebox{-18.5000124}{\makebox(0,0)[lb]{\smash{\small Intima}}}}%
    \put(0.63017867,0.23253361){\rotatebox{-18.5000124}{\makebox(0,0)[lb]{\smash{\small Media}}}}%
    \put(0.68117867,0.07253361){\rotatebox{-18.5000124}{\makebox(0,0)[lb]{\smash{\small Adventitia}}}}%
    \put(-0.02856299,0.24188722){\makebox(0,0)[rb]{\smash{\footnotesize Endothelial cell}}}%
    \put(-0.02856299,0.31688722){\makebox(0,0)[rb]{\smash{\footnotesize Internal elastic lamina}}}%
    \put(-0.02856299,0.38988722){\makebox(0,0)[rb]{\smash{\footnotesize Smooth muscle cell}}}%
    \put(-0.02856299,0.46188722){\makebox(0,0)[rb]{\smash{\footnotesize Collagen fibrils}}}%
    \put(-0.02856299,0.54188722){\makebox(0,0)[rb]{\smash{\footnotesize Elastic fibrils}}}%
    \put(-0.02856299,0.62188722){\makebox(0,0)[rb]{\smash{\footnotesize Elastic lamina}}}%
    \put(-0.02856299,0.69788722){\makebox(0,0)[rb]{\smash{\footnotesize External elastic lamina}}}%
    \put(-0.02856299,0.77188722){\makebox(0,0)[rb]{\smash{\footnotesize Bundles of collagen fibrils}}}%
    \put(0.,1.11188722){\makebox(0,0)[rb]{\smash{\footnotesize\begin{tabular}{p{3.5cm}}Helically arranged fiber- reinforced medial layers\end{tabular}}}}%
    \put(0.,1.33188722){\makebox(0,0)[rb]{\smash{\footnotesize\begin{tabular}{p{3.7cm}}Composite reinforced by collagen fibers arranged in helical structures\end{tabular}}}}%
  \end{picture}%
\endgroup%

%% file: augustin_domain.tex
\begin{picture}(0,0)%
\includegraphics{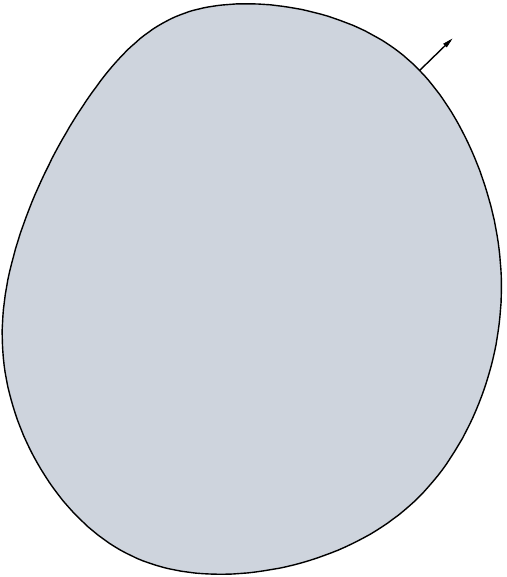}%
\end{picture}%
\setlength{\unitlength}{3522sp}%
\begingroup\makeatletter\ifx\SetFigFont\undefined%
\gdef\SetFigFont#1#2#3#4#5{%
  \reset@font\fontsize{#1}{#2pt}%
  \fontfamily{#3}\fontseries{#4}\fontshape{#5}%
  \selectfont}%
\fi\endgroup%
\begin{picture}(2710,3092)(626,-3855)
\put(1666,-2986){\makebox(0,0)[lb]{\smash{$\Omega_0$}}}
\put(811, -1276){\makebox(0,0)[lb]{\smash{$\Gamma_0$}}}
\put(3106,-1006){\makebox(0,0)[lb]{\smash{\footnotesize$\mathbf{N}_0$}}}
\end{picture}%

%% file: augustin_decomposed.tex
\begin{picture}(0,0)%
\includegraphics{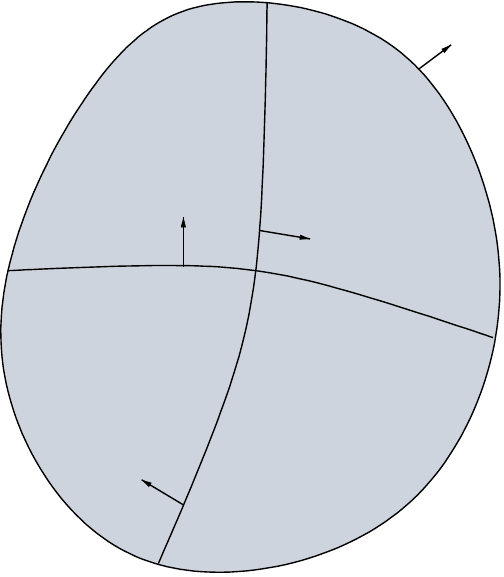}%
\end{picture}%
\setlength{\unitlength}{3522sp}%
\begingroup\makeatletter\ifx\SetFigFont\undefined%
\gdef\SetFigFont#1#2#3#4#5{%
  \reset@font\fontsize{#1}{#2pt}%
  \fontfamily{#3}\fontseries{#4}\fontshape{#5}%
  \selectfont}%
\fi\endgroup%
\begin{picture}(2710,3092)(626,-3855)
\put(1171,-1636){\makebox(0,0)[lb]{\smash{$\Omega_{0,1}$}}}
\put(2656,-1726){\makebox(0,0)[lb]{\smash{$\Omega_{0,2}$}}}
\put(2071,-3526){\makebox(0,0)[lb]{\smash{$\Omega_{0,4}$}}}
\put(766, -2851){\makebox(0,0)[lb]{\smash{$\Omega_{0,3}$}}}
\put(2116,-1276){\makebox(0,0)[lb]{\smash{$\Gamma_{0,12}$}}}
\put(2476,-2536){\makebox(0,0)[lb]{\smash{$\Gamma_{0,24}$}}}
\put(1936,-2896){\makebox(0,0)[lb]{\smash{$\Gamma_{0,34}$}}}
\put(901, -2131){\makebox(0,0)[lb]{\smash{$\Gamma_{0,13}$}}}
\put(3074,-1016){\makebox(0,0)[lb]{\smash{\footnotesize$\mathbf{N}_{0,2}$}}}
\put(2306,-1976){\makebox(0,0)[lb]{\smash{\footnotesize$\mathbf{N}_{0,1}$}}}
\put(1171,-3251){\makebox(0,0)[lb]{\smash{\footnotesize$\mathbf{N}_{0,4}$}}}
\put(1616,-1896){\makebox(0,0)[lb]{\smash{\footnotesize$\mathbf{N}_{0,3}$}}}
\put(1061,-1006){\makebox(0,0)[lb]{\smash{$\Gamma_0$}}}
\end{picture}%

%% file: augustin_classical.tex
\begingroup%
  \makeatletter%
  \providecommand\color[2][]{%
    \errmessage{(Inkscape) Color is used for the text in Inkscape, but the package 'color.sty' is not loaded}%
    \renewcommand\color[2][]{}%
  }%
  \providecommand\transparent[1]{%
    \errmessage{(Inkscape) Transparency is used (non-zero) for the text in Inkscape, but the package 'transparent.sty' is not loaded}%
    \renewcommand\transparent[1]{}%
  }%
  \providecommand\rotatebox[2]{#2}%
  \ifx\svgwidth\undefined%
    \setlength{\unitlength}{174.94763184bp}%
    \ifx\svgscale\undefined%
      \relax%
    \else%
      \setlength{\unitlength}{\unitlength * \real{\svgscale}}%
    \fi%
  \else%
    \setlength{\unitlength}{\svgwidth}%
  \fi%
  \global\let\svgwidth\undefined%
  \global\let\svgscale\undefined%
  \makeatother%
  \begin{picture}(1,0.84097737)%
    \put(0,0){\includegraphics[width=0.7\textwidth]{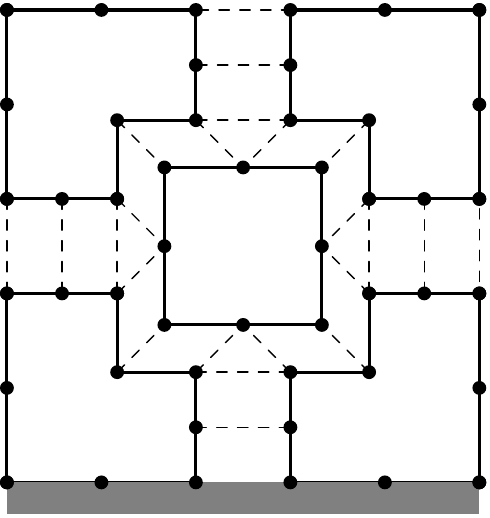}}%
    \put(0.3643716,0.4285152){\color[rgb]{0,0,0}\makebox(0,0)[lb]{\smash{$\Omega_{0,5}$}}}%
    \put(0.08664629,0.13858034){\color[rgb]{0,0,0}\makebox(0,0)[lb]{\smash{$\Omega_{0,3}$}}}%
    \put(0.61048612,0.72300026){\color[rgb]{0,0,0}\makebox(0,0)[lb]{\smash{$\Omega_{0,2}$}}}%
    \put(0.61048612,0.13858034){\color[rgb]{0,0,0}\makebox(0,0)[lb]{\smash{$\Omega_{0,4}$}}}%
    \put(0.08664629,0.72300026){\color[rgb]{0,0,0}\makebox(0,0)[lb]{\smash{$\Omega_{0,1}$}}}%
  \end{picture}%
\endgroup%

%% file: augustin_allfloating.tex
\begingroup%
  \makeatletter%
  \providecommand\color[2][]{%
    \errmessage{(Inkscape) Color is used for the text in Inkscape, but the package 'color.sty' is not loaded}%
    \renewcommand\color[2][]{}%
  }%
  \providecommand\transparent[1]{%
    \errmessage{(Inkscape) Transparency is used (non-zero) for the text in Inkscape, but the package 'transparent.sty' is not loaded}%
    \renewcommand\transparent[1]{}%
  }%
  \providecommand\rotatebox[2]{#2}%
  \ifx\svgwidth\undefined%
    \setlength{\unitlength}{174.95539856bp}%
    \ifx\svgscale\undefined%
      \relax%
    \else%
      \setlength{\unitlength}{\unitlength * \real{\svgscale}}%
    \fi%
  \else%
    \setlength{\unitlength}{\svgwidth}%
  \fi%
  \global\let\svgwidth\undefined%
  \global\let\svgscale\undefined%
  \makeatother%
  \begin{picture}(1,0.90574513)%
    \put(0,0){\includegraphics[width=0.7\textwidth]{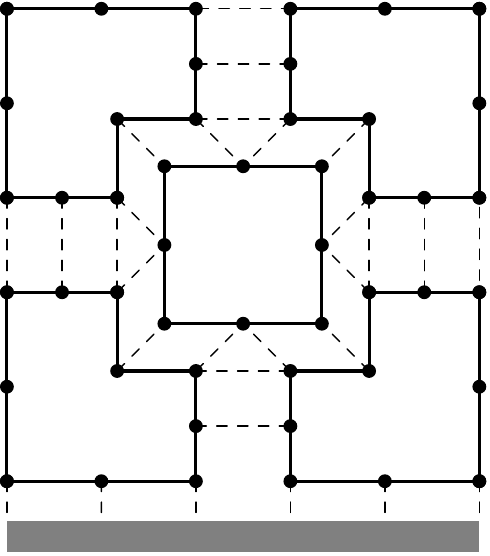}}%
    \put(0.36439984,0.49335123){\color[rgb]{0,0,0}\makebox(0,0)[lb]{\smash{$\Omega_{0,5}$}}}%
    \put(0.08668688,0.20342921){\color[rgb]{0,0,0}\makebox(0,0)[lb]{\smash{$\Omega_{0,3}$}}}%
    \put(0.61050341,0.78782321){\color[rgb]{0,0,0}\makebox(0,0)[lb]{\smash{$\Omega_{0,2}$}}}%
    \put(0.61050341,0.20342921){\color[rgb]{0,0,0}\makebox(0,0)[lb]{\smash{$\Omega_{0,4}$}}}%
    \put(0.08668688,0.78782321){\color[rgb]{0,0,0}\makebox(0,0)[lb]{\smash{$\Omega_{0,1}$}}}%
  \end{picture}%
\endgroup%

%% file: augustin_aorta_points.tex
\begingroup%
  \makeatletter%
  \providecommand\color[2][]{%
    \errmessage{(Inkscape) Color is used for the text in Inkscape, but the package 'color.sty' is not loaded}%
    \renewcommand\color[2][]{}%
  }%
  \providecommand\transparent[1]{%
    \errmessage{(Inkscape) Transparency is used (non-zero) for the text in Inkscape, but the package 'transparent.sty' is not loaded}%
    \renewcommand\transparent[1]{}%
  }%
  \providecommand\rotatebox[2]{#2}%
  \ifx\svgwidth\undefined%
    \setlength{\unitlength}{1524.8bp}%
    \ifx\svgscale\undefined%
      \relax%
    \else%
      \setlength{\unitlength}{\unitlength * \real{\svgscale}}%
    \fi%
  \else%
    \setlength{\unitlength}{\svgwidth}%
  \fi%
  \global\let\svgwidth\undefined%
  \global\let\svgscale\undefined%
  \makeatother%
  \begin{picture}(1,0.87303253)%
    \put(0,0){\includegraphics[width=\unitlength]{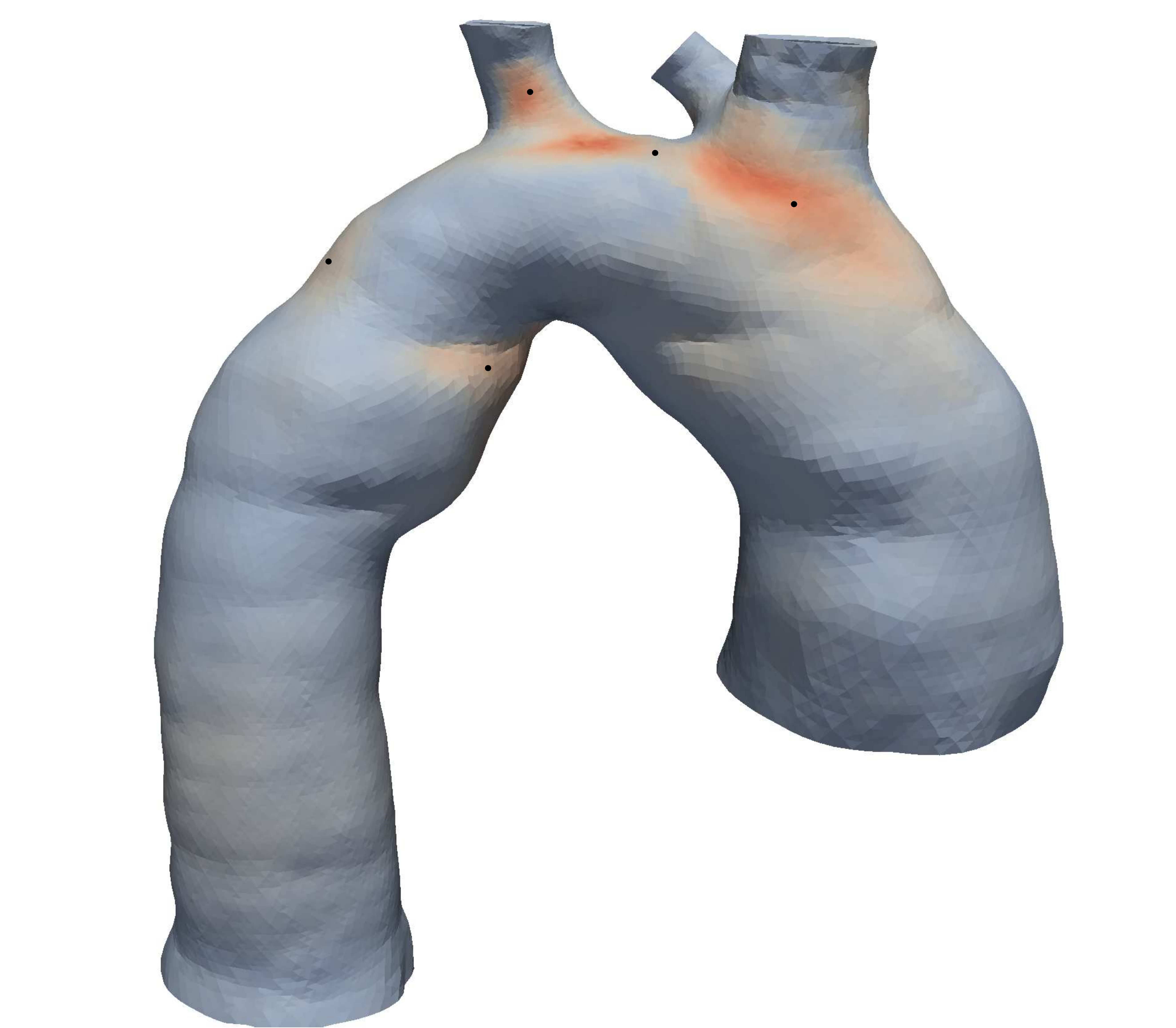}}%
    \put(0.286695,0.64737582){\color[rgb]{0,0,0}\makebox(0,0)[lb]{\smash{A}}}%
    \put(0.45778718,0.79164814){\color[rgb]{0,0,0}\makebox(0,0)[lb]{\smash{B}}}%
    \put(0.38183542,0.55443113){\color[rgb]{0,0,0}\makebox(0,0)[lb]{\smash{C}}}%
    \put(0.56414174,0.73615879){\color[rgb]{0,0,0}\makebox(0,0)[lb]{\smash{D}}}%
    \put(0.68205673,0.68945524){\color[rgb]{0,0,0}\makebox(0,0)[lb]{\smash{E}}}%
  \end{picture}%
\endgroup%

%% file: augustin_stress_vs_displacement.tex
\begingroup
  \makeatletter
  \providecommand\color[2][]{%
    \GenericError{(gnuplot) \space\space\space\@spaces}{%
      Package color not loaded in conjunction with
      terminal option `colourtext'%
    }{See the gnuplot documentation for explanation.%
    }{Either use 'blacktext' in gnuplot or load the package
      color.sty in LaTeX.}%
    \renewcommand\color[2][]{}%
  }%
  \providecommand\includegraphics[2][]{%
    \GenericError{(gnuplot) \space\space\space\@spaces}{%
      Package graphicx or graphics not loaded%
    }{See the gnuplot documentation for explanation.%
    }{The gnuplot epslatex terminal needs graphicx.sty or graphics.sty.}%
    \renewcommand\includegraphics[2][]{}%
  }%
  \providecommand\rotatebox[2]{#2}%
  \@ifundefined{ifGPcolor}{%
    \newif\ifGPcolor
    \GPcolortrue
  }{}%
  \@ifundefined{ifGPblacktext}{%
    \newif\ifGPblacktext
    \GPblacktexttrue
  }{}%
  \let\gplgaddtomacro\g@addto@macro
  \gdef\gplbacktext{}%
  \gdef\gplfronttext{}%
  \makeatother
  \ifGPblacktext
    \def\colorrgb#1{}%
    \def\colorgray#1{}%
  \else
    \ifGPcolor
      \def\colorrgb#1{\color[rgb]{#1}}%
      \def\colorgray#1{\color[gray]{#1}}%
      \expandafter\def\csname LTw\endcsname{\color{white}}%
      \expandafter\def\csname LTb\endcsname{\color{black}}%
      \expandafter\def\csname LTa\endcsname{\color{black}}%
      \expandafter\def\csname LT0\endcsname{\color[rgb]{1,0,0}}%
      \expandafter\def\csname LT1\endcsname{\color[rgb]{0,1,0}}%
      \expandafter\def\csname LT2\endcsname{\color[rgb]{0,0,1}}%
      \expandafter\def\csname LT3\endcsname{\color[rgb]{1,0,1}}%
      \expandafter\def\csname LT4\endcsname{\color[rgb]{0,1,1}}%
      \expandafter\def\csname LT5\endcsname{\color[rgb]{1,1,0}}%
      \expandafter\def\csname LT6\endcsname{\color[rgb]{0,0,0}}%
      \expandafter\def\csname LT7\endcsname{\color[rgb]{1,0.3,0}}%
      \expandafter\def\csname LT8\endcsname{\color[rgb]{0.5,0.5,0.5}}%
    \else
      \def\colorrgb#1{\color{black}}%
      \def\colorgray#1{\color[gray]{#1}}%
      \expandafter\def\csname LTw\endcsname{\color{white}}%
      \expandafter\def\csname LTb\endcsname{\color{black}}%
      \expandafter\def\csname LTa\endcsname{\color{black}}%
      \expandafter\def\csname LT0\endcsname{\color{black}}%
      \expandafter\def\csname LT1\endcsname{\color{black}}%
      \expandafter\def\csname LT2\endcsname{\color{black}}%
      \expandafter\def\csname LT3\endcsname{\color{black}}%
      \expandafter\def\csname LT4\endcsname{\color{black}}%
      \expandafter\def\csname LT5\endcsname{\color{black}}%
      \expandafter\def\csname LT6\endcsname{\color{black}}%
      \expandafter\def\csname LT7\endcsname{\color{black}}%
      \expandafter\def\csname LT8\endcsname{\color{black}}%
    \fi
  \fi
  \setlength{\unitlength}{0.0500bp}%
  \begin{picture}(4464.00,3124.80)%
    \gplgaddtomacro\gplbacktext{%
      \csname LTb\endcsname%
      \put(924,484){\makebox(0,0)[r]{\strut{}\tiny 0}}%
      \csname LTb\endcsname%
      \put(924,959){\makebox(0,0)[r]{\strut{}\tiny 100}}%
      \csname LTb\endcsname%
      \put(924,1434){\makebox(0,0)[r]{\strut{}\tiny 200}}%
      \csname LTb\endcsname%
      \put(924,1910){\makebox(0,0)[r]{\strut{}\tiny 300}}%
      \csname LTb\endcsname%
      \put(924,2385){\makebox(0,0)[r]{\strut{}\tiny 400}}%
      \csname LTb\endcsname%
      \put(924,2860){\makebox(0,0)[r]{\strut{}\tiny 500}}%
      \put(2525,352){\makebox(0,0){\strut{}\tiny 0.5}}%
      \put(4067,352){\makebox(0,0){\strut{}\tiny 1}}%
      \put(418,1672){\rotatebox{-270}{\makebox(0,0){\strut{}{\tiny Stress magnitude} $\scriptstyle \sigma_\mathrm{mag}$ {\tiny [kPa]} }}}%
      \put(2528,154){\makebox(0,0){\strut{}{\tiny Relative displacement} $\scriptstyle u_\mathrm{rel}$ {\tiny [-]}}}%
    }%
    \gplgaddtomacro\gplfronttext{%
      \csname LTb\endcsname%
      \put(1518,2742){\makebox(0,0)[r]{\strut{}\tiny A}}%
      \csname LTb\endcsname%
      \put(1518,2632){\makebox(0,0)[r]{\strut{}\tiny B}}%
      \csname LTb\endcsname%
      \put(1518,2522){\makebox(0,0)[r]{\strut{}\tiny C}}%
      \csname LTb\endcsname%
      \put(1518,2412){\makebox(0,0)[r]{\strut{}\tiny D}}%
      \csname LTb\endcsname%
      \put(1518,2302){\makebox(0,0)[r]{\strut{}\tiny E}}%
    }%
    \gplbacktext
    \put(0,0){\includegraphics{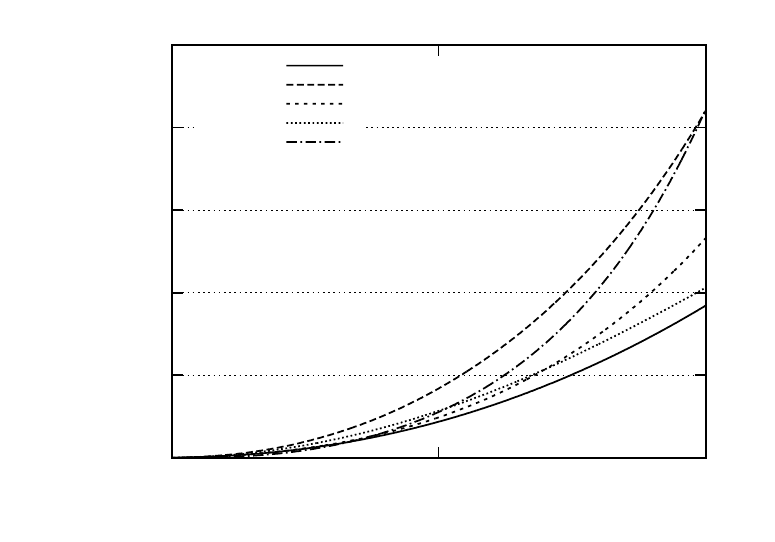}}%
    \gplfronttext
  \end{picture}%
\endgroup

%% file: augustin_displacement_vs_pressure.tex
\begingroup
  \makeatletter
  \providecommand\color[2][]{%
    \GenericError{(gnuplot) \space\space\space\@spaces}{%
      Package color not loaded in conjunction with
      terminal option `colourtext'%
    }{See the gnuplot documentation for explanation.%
    }{Either use 'blacktext' in gnuplot or load the package
      color.sty in LaTeX.}%
    \renewcommand\color[2][]{}%
  }%
  \providecommand\includegraphics[2][]{%
    \GenericError{(gnuplot) \space\space\space\@spaces}{%
      Package graphicx or graphics not loaded%
    }{See the gnuplot documentation for explanation.%
    }{The gnuplot epslatex terminal needs graphicx.sty or graphics.sty.}%
    \renewcommand\includegraphics[2][]{}%
  }%
  \providecommand\rotatebox[2]{#2}%
  \@ifundefined{ifGPcolor}{%
    \newif\ifGPcolor
    \GPcolortrue
  }{}%
  \@ifundefined{ifGPblacktext}{%
    \newif\ifGPblacktext
    \GPblacktexttrue
  }{}%
  \let\gplgaddtomacro\g@addto@macro
  \gdef\gplbacktext{}%
  \gdef\gplfronttext{}%
  \makeatother
  \ifGPblacktext
    \def\colorrgb#1{}%
    \def\colorgray#1{}%
  \else
    \ifGPcolor
      \def\colorrgb#1{\color[rgb]{#1}}%
      \def\colorgray#1{\color[gray]{#1}}%
      \expandafter\def\csname LTw\endcsname{\color{white}}%
      \expandafter\def\csname LTb\endcsname{\color{black}}%
      \expandafter\def\csname LTa\endcsname{\color{black}}%
      \expandafter\def\csname LT0\endcsname{\color[rgb]{1,0,0}}%
      \expandafter\def\csname LT1\endcsname{\color[rgb]{0,1,0}}%
      \expandafter\def\csname LT2\endcsname{\color[rgb]{0,0,1}}%
      \expandafter\def\csname LT3\endcsname{\color[rgb]{1,0,1}}%
      \expandafter\def\csname LT4\endcsname{\color[rgb]{0,1,1}}%
      \expandafter\def\csname LT5\endcsname{\color[rgb]{1,1,0}}%
      \expandafter\def\csname LT6\endcsname{\color[rgb]{0,0,0}}%
      \expandafter\def\csname LT7\endcsname{\color[rgb]{1,0.3,0}}%
      \expandafter\def\csname LT8\endcsname{\color[rgb]{0.5,0.5,0.5}}%
    \else
      \def\colorrgb#1{\color{black}}%
      \def\colorgray#1{\color[gray]{#1}}%
      \expandafter\def\csname LTw\endcsname{\color{white}}%
      \expandafter\def\csname LTb\endcsname{\color{black}}%
      \expandafter\def\csname LTa\endcsname{\color{black}}%
      \expandafter\def\csname LT0\endcsname{\color{black}}%
      \expandafter\def\csname LT1\endcsname{\color{black}}%
      \expandafter\def\csname LT2\endcsname{\color{black}}%
      \expandafter\def\csname LT3\endcsname{\color{black}}%
      \expandafter\def\csname LT4\endcsname{\color{black}}%
      \expandafter\def\csname LT5\endcsname{\color{black}}%
      \expandafter\def\csname LT6\endcsname{\color{black}}%
      \expandafter\def\csname LT7\endcsname{\color{black}}%
      \expandafter\def\csname LT8\endcsname{\color{black}}%
    \fi
  \fi
  \setlength{\unitlength}{0.0500bp}%
  \begin{picture}(4464.00,3124.80)%
    \gplgaddtomacro\gplbacktext{%
      \csname LTb\endcsname%
      \put(924,484){\makebox(0,0)[r]{\strut{}\tiny 0}}%
      \csname LTb\endcsname%
      \put(924,880){\makebox(0,0)[r]{\strut{}\tiny 50}}%
      \csname LTb\endcsname%
      \put(924,1276){\makebox(0,0)[r]{\strut{}\tiny 100}}%
      \csname LTb\endcsname%
      \put(924,1672){\makebox(0,0)[r]{\strut{}\tiny 150}}%
      \csname LTb\endcsname%
      \put(924,2068){\makebox(0,0)[r]{\strut{}\tiny 200}}%
      \csname LTb\endcsname%
      \put(924,2464){\makebox(0,0)[r]{\strut{}\tiny 250}}%
      \csname LTb\endcsname%
      \put(924,2860){\makebox(0,0)[r]{\strut{}\tiny 300}}%
      \put(990,352){\makebox(0,0){\strut{}\tiny 0}}%
      \put(1759,352){\makebox(0,0){\strut{}\tiny 0.5}}%
      \put(2529,352){\makebox(0,0){\strut{}\tiny 1}}%
      \put(3298,352){\makebox(0,0){\strut{}\tiny 1.5}}%
      \put(4067,352){\makebox(0,0){\strut{}\tiny 2}}%
      \put(418,1672){\rotatebox{-270}{\makebox(0,0){\strut{}{\tiny Internal pressure} $\scriptstyle p$ {\tiny [mmHg]}}}}%
      \put(2528,154){\makebox(0,0){\strut{}{\tiny Displacement norm} $\scriptstyle u_\mathrm{norm}$ {\tiny [mm]}}}%
    }%
    \gplgaddtomacro\gplfronttext{%
      \csname LTb\endcsname%
      \put(3344,1042){\makebox(0,0)[r]{\strut{}\tiny A}}%
      \csname LTb\endcsname%
      \put(3344,932){\makebox(0,0)[r]{\strut{}\tiny B}}%
      \csname LTb\endcsname%
      \put(3344,822){\makebox(0,0)[r]{\strut{}\tiny C}}%
      \csname LTb\endcsname%
      \put(3344,712){\makebox(0,0)[r]{\strut{}\tiny D}}%
      \csname LTb\endcsname%
      \put(3344,602){\makebox(0,0)[r]{\strut{}\tiny E}}%
    }%
    \gplbacktext
    \put(0,0){\includegraphics{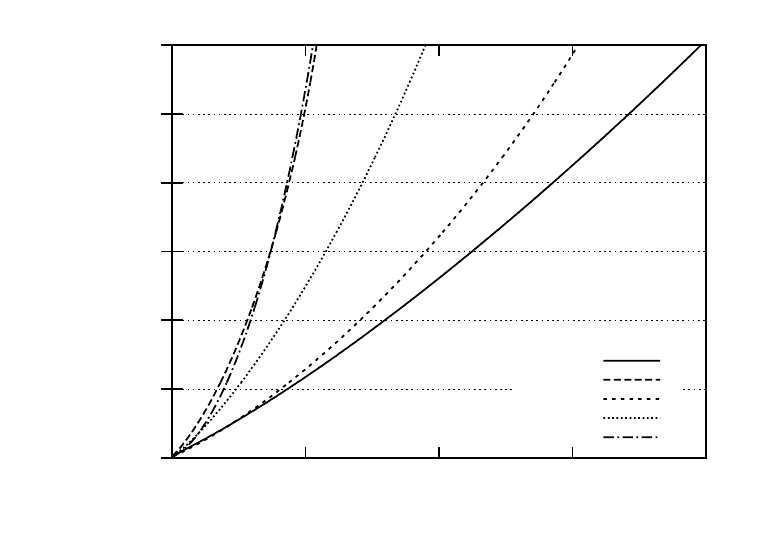}}%
    \gplfronttext
  \end{picture}%
\endgroup

%% file: augustin_timestepping_iterations.tex
\begingroup
  \makeatletter
  \providecommand\color[2][]{%
    \GenericError{(gnuplot) \space\space\space\@spaces}{%
      Package color not loaded in conjunction with
      terminal option `colourtext'%
    }{See the gnuplot documentation for explanation.%
    }{Either use 'blacktext' in gnuplot or load the package
      color.sty in LaTeX.}%
    \renewcommand\color[2][]{}%
  }%
  \providecommand\includegraphics[2][]{%
    \GenericError{(gnuplot) \space\space\space\@spaces}{%
      Package graphicx or graphics not loaded%
    }{See the gnuplot documentation for explanation.%
    }{The gnuplot epslatex terminal needs graphicx.sty or graphics.sty.}%
    \renewcommand\includegraphics[2][]{}%
  }%
  \providecommand\rotatebox[2]{#2}%
  \@ifundefined{ifGPcolor}{%
    \newif\ifGPcolor
    \GPcolortrue
  }{}%
  \@ifundefined{ifGPblacktext}{%
    \newif\ifGPblacktext
    \GPblacktexttrue
  }{}%
  \let\gplgaddtomacro\g@addto@macro
  \gdef\gplbacktext{}%
  \gdef\gplfronttext{}%
  \makeatother
  \ifGPblacktext
    \def\colorrgb#1{}%
    \def\colorgray#1{}%
  \else
    \ifGPcolor
      \def\colorrgb#1{\color[rgb]{#1}}%
      \def\colorgray#1{\color[gray]{#1}}%
      \expandafter\def\csname LTw\endcsname{\color{white}}%
      \expandafter\def\csname LTb\endcsname{\color{black}}%
      \expandafter\def\csname LTa\endcsname{\color{black}}%
      \expandafter\def\csname LT0\endcsname{\color[rgb]{1,0,0}}%
      \expandafter\def\csname LT1\endcsname{\color[rgb]{0,1,0}}%
      \expandafter\def\csname LT2\endcsname{\color[rgb]{0,0,1}}%
      \expandafter\def\csname LT3\endcsname{\color[rgb]{1,0,1}}%
      \expandafter\def\csname LT4\endcsname{\color[rgb]{0,1,1}}%
      \expandafter\def\csname LT5\endcsname{\color[rgb]{1,1,0}}%
      \expandafter\def\csname LT6\endcsname{\color[rgb]{0,0,0}}%
      \expandafter\def\csname LT7\endcsname{\color[rgb]{1,0.3,0}}%
      \expandafter\def\csname LT8\endcsname{\color[rgb]{0.5,0.5,0.5}}%
    \else
      \def\colorrgb#1{\color{black}}%
      \def\colorgray#1{\color[gray]{#1}}%
      \expandafter\def\csname LTw\endcsname{\color{white}}%
      \expandafter\def\csname LTb\endcsname{\color{black}}%
      \expandafter\def\csname LTa\endcsname{\color{black}}%
      \expandafter\def\csname LT0\endcsname{\color{black}}%
      \expandafter\def\csname LT1\endcsname{\color{black}}%
      \expandafter\def\csname LT2\endcsname{\color{black}}%
      \expandafter\def\csname LT3\endcsname{\color{black}}%
      \expandafter\def\csname LT4\endcsname{\color{black}}%
      \expandafter\def\csname LT5\endcsname{\color{black}}%
      \expandafter\def\csname LT6\endcsname{\color{black}}%
      \expandafter\def\csname LT7\endcsname{\color{black}}%
      \expandafter\def\csname LT8\endcsname{\color{black}}%
    \fi
  \fi
  \setlength{\unitlength}{0.0500bp}%
  \begin{picture}(4464.00,3124.80)%
    \gplgaddtomacro\gplbacktext{%
      \csname LTb\endcsname%
      \put(924,484){\makebox(0,0)[r]{\strut{}\tiny 200}}%
      \csname LTb\endcsname%
      \put(924,1078){\makebox(0,0)[r]{\strut{}\tiny 250}}%
      \csname LTb\endcsname%
      \put(924,1672){\makebox(0,0)[r]{\strut{}\tiny 300}}%
      \csname LTb\endcsname%
      \put(924,2266){\makebox(0,0)[r]{\strut{}\tiny 350}}%
      \csname LTb\endcsname%
      \put(924,2860){\makebox(0,0)[r]{\strut{}\tiny 400}}%
      \put(990,352){\makebox(0,0){\strut{}\tiny 0}}%
      \put(1529,352){\makebox(0,0){\strut{}\tiny 100}}%
      \put(2068,352){\makebox(0,0){\strut{}\tiny 200}}%
      \put(2607,352){\makebox(0,0){\strut{}\tiny 300}}%
      \put(3146,352){\makebox(0,0){\strut{}\tiny 400}}%
      \put(3684,352){\makebox(0,0){\strut{}\tiny 500}}%
      \put(418,1672){\rotatebox{-270}{\makebox(0,0){\strut{}\tiny Iterations [-]}}}%
      \put(2528,154){\makebox(0,0){\strut{}\tiny Time step [-]}}%
    }%
    \gplgaddtomacro\gplfronttext{%
    }%
    \gplbacktext
    \put(0,0){\includegraphics{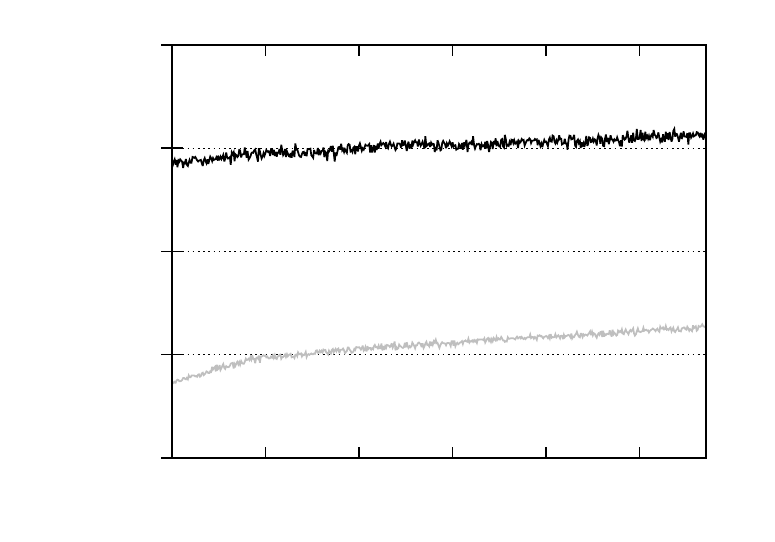}}%
    \gplfronttext
  \end{picture}%
\endgroup

%% file: augustin_timestepping_solve_time.tex
\begingroup
  \makeatletter
  \providecommand\color[2][]{%
    \GenericError{(gnuplot) \space\space\space\@spaces}{%
      Package color not loaded in conjunction with
      terminal option `colourtext'%
    }{See the gnuplot documentation for explanation.%
    }{Either use 'blacktext' in gnuplot or load the package
      color.sty in LaTeX.}%
    \renewcommand\color[2][]{}%
  }%
  \providecommand\includegraphics[2][]{%
    \GenericError{(gnuplot) \space\space\space\@spaces}{%
      Package graphicx or graphics not loaded%
    }{See the gnuplot documentation for explanation.%
    }{The gnuplot epslatex terminal needs graphicx.sty or graphics.sty.}%
    \renewcommand\includegraphics[2][]{}%
  }%
  \providecommand\rotatebox[2]{#2}%
  \@ifundefined{ifGPcolor}{%
    \newif\ifGPcolor
    \GPcolortrue
  }{}%
  \@ifundefined{ifGPblacktext}{%
    \newif\ifGPblacktext
    \GPblacktexttrue
  }{}%
  \let\gplgaddtomacro\g@addto@macro
  \gdef\gplbacktext{}%
  \gdef\gplfronttext{}%
  \makeatother
  \ifGPblacktext
    \def\colorrgb#1{}%
    \def\colorgray#1{}%
  \else
    \ifGPcolor
      \def\colorrgb#1{\color[rgb]{#1}}%
      \def\colorgray#1{\color[gray]{#1}}%
      \expandafter\def\csname LTw\endcsname{\color{white}}%
      \expandafter\def\csname LTb\endcsname{\color{black}}%
      \expandafter\def\csname LTa\endcsname{\color{black}}%
      \expandafter\def\csname LT0\endcsname{\color[rgb]{1,0,0}}%
      \expandafter\def\csname LT1\endcsname{\color[rgb]{0,1,0}}%
      \expandafter\def\csname LT2\endcsname{\color[rgb]{0,0,1}}%
      \expandafter\def\csname LT3\endcsname{\color[rgb]{1,0,1}}%
      \expandafter\def\csname LT4\endcsname{\color[rgb]{0,1,1}}%
      \expandafter\def\csname LT5\endcsname{\color[rgb]{1,1,0}}%
      \expandafter\def\csname LT6\endcsname{\color[rgb]{0,0,0}}%
      \expandafter\def\csname LT7\endcsname{\color[rgb]{1,0.3,0}}%
      \expandafter\def\csname LT8\endcsname{\color[rgb]{0.5,0.5,0.5}}%
    \else
      \def\colorrgb#1{\color{black}}%
      \def\colorgray#1{\color[gray]{#1}}%
      \expandafter\def\csname LTw\endcsname{\color{white}}%
      \expandafter\def\csname LTb\endcsname{\color{black}}%
      \expandafter\def\csname LTa\endcsname{\color{black}}%
      \expandafter\def\csname LT0\endcsname{\color{black}}%
      \expandafter\def\csname LT1\endcsname{\color{black}}%
      \expandafter\def\csname LT2\endcsname{\color{black}}%
      \expandafter\def\csname LT3\endcsname{\color{black}}%
      \expandafter\def\csname LT4\endcsname{\color{black}}%
      \expandafter\def\csname LT5\endcsname{\color{black}}%
      \expandafter\def\csname LT6\endcsname{\color{black}}%
      \expandafter\def\csname LT7\endcsname{\color{black}}%
      \expandafter\def\csname LT8\endcsname{\color{black}}%
    \fi
  \fi
  \setlength{\unitlength}{0.0500bp}%
  \begin{picture}(4464.00,3124.80)%
    \gplgaddtomacro\gplbacktext{%
      \csname LTb\endcsname%
      \put(858,810){\makebox(0,0)[r]{\strut{}\tiny 250}}%
      \csname LTb\endcsname%
      \put(858,1350){\makebox(0,0)[r]{\strut{}\tiny 300}}%
      \csname LTb\endcsname%
      \put(858,1890){\makebox(0,0)[r]{\strut{}\tiny 350}}%
      \csname LTb\endcsname%
      \put(858,2430){\makebox(0,0)[r]{\strut{}\tiny 400}}%
      \csname LTb\endcsname%
      \put(858,2970){\makebox(0,0)[r]{\strut{}\tiny 450}}%
      \put(990,352){\makebox(0,0){\strut{}\tiny 0}}%
      \put(1529,352){\makebox(0,0){\strut{}\tiny 100}}%
      \put(2068,352){\makebox(0,0){\strut{}\tiny 200}}%
      \put(2607,352){\makebox(0,0){\strut{}\tiny 300}}%
      \put(3146,352){\makebox(0,0){\strut{}\tiny 400}}%
      \put(3684,352){\makebox(0,0){\strut{}\tiny 500}}%
      \put(418,1672){\rotatebox{-270}{\makebox(0,0){\strut{}\tiny Time [s]}}}%
      \put(2528,154){\makebox(0,0){\strut{}\tiny Time step [-]}}%
    }%
    \gplgaddtomacro\gplfronttext{%
    }%
    \gplbacktext
    \put(0,0){\includegraphics{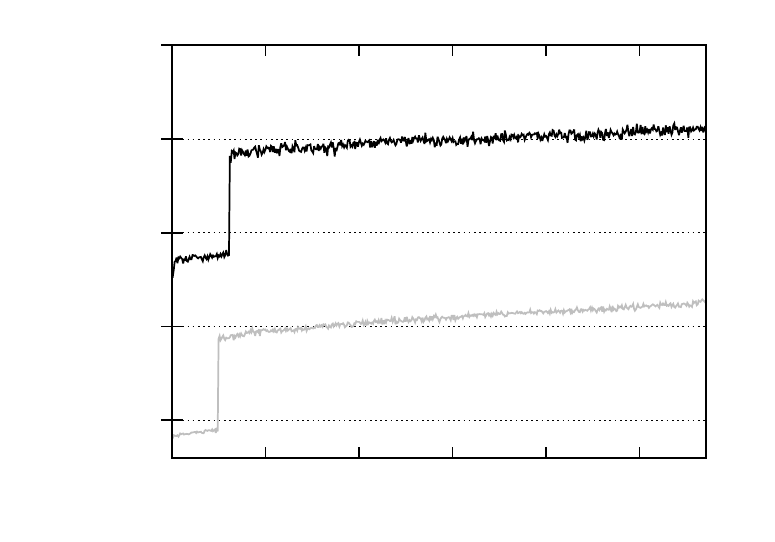}}%
    \gplfronttext
  \end{picture}%
\endgroup

%% file: augustin_solve_time_aorta.tex
\begingroup
  \makeatletter
  \providecommand\color[2][]{%
    \GenericError{(gnuplot) \space\space\space\@spaces}{%
      Package color not loaded in conjunction with
      terminal option `colourtext'%
    }{See the gnuplot documentation for explanation.%
    }{Either use 'blacktext' in gnuplot or load the package
      color.sty in LaTeX.}%
    \renewcommand\color[2][]{}%
  }%
  \providecommand\includegraphics[2][]{%
    \GenericError{(gnuplot) \space\space\space\@spaces}{%
      Package graphicx or graphics not loaded%
    }{See the gnuplot documentation for explanation.%
    }{The gnuplot epslatex terminal needs graphicx.sty or graphics.sty.}%
    \renewcommand\includegraphics[2][]{}%
  }%
  \providecommand\rotatebox[2]{#2}%
  \@ifundefined{ifGPcolor}{%
    \newif\ifGPcolor
    \GPcolortrue
  }{}%
  \@ifundefined{ifGPblacktext}{%
    \newif\ifGPblacktext
    \GPblacktexttrue
  }{}%
  \let\gplgaddtomacro\g@addto@macro
  \gdef\gplbacktext{}%
  \gdef\gplfronttext{}%
  \makeatother
  \ifGPblacktext
    \def\colorrgb#1{}%
    \def\colorgray#1{}%
  \else
    \ifGPcolor
      \def\colorrgb#1{\color[rgb]{#1}}%
      \def\colorgray#1{\color[gray]{#1}}%
      \expandafter\def\csname LTw\endcsname{\color{white}}%
      \expandafter\def\csname LTb\endcsname{\color{black}}%
      \expandafter\def\csname LTa\endcsname{\color{black}}%
      \expandafter\def\csname LT0\endcsname{\color[rgb]{1,0,0}}%
      \expandafter\def\csname LT1\endcsname{\color[rgb]{0,1,0}}%
      \expandafter\def\csname LT2\endcsname{\color[rgb]{0,0,1}}%
      \expandafter\def\csname LT3\endcsname{\color[rgb]{1,0,1}}%
      \expandafter\def\csname LT4\endcsname{\color[rgb]{0,1,1}}%
      \expandafter\def\csname LT5\endcsname{\color[rgb]{1,1,0}}%
      \expandafter\def\csname LT6\endcsname{\color[rgb]{0,0,0}}%
      \expandafter\def\csname LT7\endcsname{\color[rgb]{1,0.3,0}}%
      \expandafter\def\csname LT8\endcsname{\color[rgb]{0.5,0.5,0.5}}%
    \else
      \def\colorrgb#1{\color{black}}%
      \def\colorgray#1{\color[gray]{#1}}%
      \expandafter\def\csname LTw\endcsname{\color{white}}%
      \expandafter\def\csname LTb\endcsname{\color{black}}%
      \expandafter\def\csname LTa\endcsname{\color{black}}%
      \expandafter\def\csname LT0\endcsname{\color{black}}%
      \expandafter\def\csname LT1\endcsname{\color{black}}%
      \expandafter\def\csname LT2\endcsname{\color{black}}%
      \expandafter\def\csname LT3\endcsname{\color{black}}%
      \expandafter\def\csname LT4\endcsname{\color{black}}%
      \expandafter\def\csname LT5\endcsname{\color{black}}%
      \expandafter\def\csname LT6\endcsname{\color{black}}%
      \expandafter\def\csname LT7\endcsname{\color{black}}%
      \expandafter\def\csname LT8\endcsname{\color{black}}%
    \fi
  \fi
  \setlength{\unitlength}{0.0500bp}%
  \begin{picture}(4680.00,3276.00)%
    \gplgaddtomacro\gplbacktext{%
      \csname LTb\endcsname%
      \put(1056,738){\makebox(0,0)[r]{\strut{}\tiny 10}}%
      \csname LTb\endcsname%
      \put(1056,1580){\makebox(0,0)[r]{\strut{}\tiny 100}}%
      \csname LTb\endcsname%
      \put(1056,2422){\makebox(0,0)[r]{\strut{}\tiny 1000}}%
      \put(1122,352){\makebox(0,0){\strut{}\tiny 16}}%
      \put(1754,352){\makebox(0,0){\strut{}\tiny 32}}%
      \put(2386,352){\makebox(0,0){\strut{}\tiny 64}}%
      \put(3019,352){\makebox(0,0){\strut{}\tiny 128}}%
      \put(3651,352){\makebox(0,0){\strut{}\tiny 256}}%
      \put(4283,352){\makebox(0,0){\strut{}\tiny 512}}%
      \put(616,1747){\rotatebox{-270}{\makebox(0,0){\strut{}\tiny Time [s]}}}%
      \put(2702,154){\makebox(0,0){\strut{}{\tiny Number of processes} $\scriptstyle P$ {\tiny [-]}}}%
    }%
    \gplgaddtomacro\gplfronttext{%
      \csname LTb\endcsname%
      \put(3560,2893){\makebox(0,0)[r]{\strut{}\tiny local time}}%
      \csname LTb\endcsname%
      \put(3560,2783){\makebox(0,0)[r]{\strut{}\tiny global CG time}}%
      \csname LTb\endcsname%
      \put(3560,2673){\makebox(0,0)[r]{\strut{}\tiny total time}}%
    }%
    \gplbacktext
    \put(0,0){\includegraphics{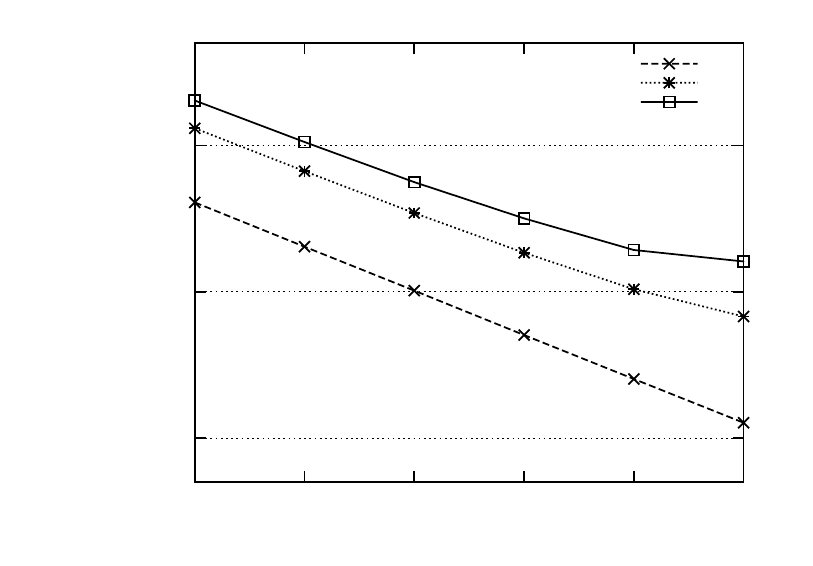}}%
    \gplfronttext
  \end{picture}%
\endgroup

%% file: augustin_solve_time_carotis.tex
\begingroup
  \makeatletter
  \providecommand\color[2][]{%
    \GenericError{(gnuplot) \space\space\space\@spaces}{%
      Package color not loaded in conjunction with
      terminal option `colourtext'%
    }{See the gnuplot documentation for explanation.%
    }{Either use 'blacktext' in gnuplot or load the package
      color.sty in LaTeX.}%
    \renewcommand\color[2][]{}%
  }%
  \providecommand\includegraphics[2][]{%
    \GenericError{(gnuplot) \space\space\space\@spaces}{%
      Package graphicx or graphics not loaded%
    }{See the gnuplot documentation for explanation.%
    }{The gnuplot epslatex terminal needs graphicx.sty or graphics.sty.}%
    \renewcommand\includegraphics[2][]{}%
  }%
  \providecommand\rotatebox[2]{#2}%
  \@ifundefined{ifGPcolor}{%
    \newif\ifGPcolor
    \GPcolortrue
  }{}%
  \@ifundefined{ifGPblacktext}{%
    \newif\ifGPblacktext
    \GPblacktexttrue
  }{}%
  \let\gplgaddtomacro\g@addto@macro
  \gdef\gplbacktext{}%
  \gdef\gplfronttext{}%
  \makeatother
  \ifGPblacktext
    \def\colorrgb#1{}%
    \def\colorgray#1{}%
  \else
    \ifGPcolor
      \def\colorrgb#1{\color[rgb]{#1}}%
      \def\colorgray#1{\color[gray]{#1}}%
      \expandafter\def\csname LTw\endcsname{\color{white}}%
      \expandafter\def\csname LTb\endcsname{\color{black}}%
      \expandafter\def\csname LTa\endcsname{\color{black}}%
      \expandafter\def\csname LT0\endcsname{\color[rgb]{1,0,0}}%
      \expandafter\def\csname LT1\endcsname{\color[rgb]{0,1,0}}%
      \expandafter\def\csname LT2\endcsname{\color[rgb]{0,0,1}}%
      \expandafter\def\csname LT3\endcsname{\color[rgb]{1,0,1}}%
      \expandafter\def\csname LT4\endcsname{\color[rgb]{0,1,1}}%
      \expandafter\def\csname LT5\endcsname{\color[rgb]{1,1,0}}%
      \expandafter\def\csname LT6\endcsname{\color[rgb]{0,0,0}}%
      \expandafter\def\csname LT7\endcsname{\color[rgb]{1,0.3,0}}%
      \expandafter\def\csname LT8\endcsname{\color[rgb]{0.5,0.5,0.5}}%
    \else
      \def\colorrgb#1{\color{black}}%
      \def\colorgray#1{\color[gray]{#1}}%
      \expandafter\def\csname LTw\endcsname{\color{white}}%
      \expandafter\def\csname LTb\endcsname{\color{black}}%
      \expandafter\def\csname LTa\endcsname{\color{black}}%
      \expandafter\def\csname LT0\endcsname{\color{black}}%
      \expandafter\def\csname LT1\endcsname{\color{black}}%
      \expandafter\def\csname LT2\endcsname{\color{black}}%
      \expandafter\def\csname LT3\endcsname{\color{black}}%
      \expandafter\def\csname LT4\endcsname{\color{black}}%
      \expandafter\def\csname LT5\endcsname{\color{black}}%
      \expandafter\def\csname LT6\endcsname{\color{black}}%
      \expandafter\def\csname LT7\endcsname{\color{black}}%
      \expandafter\def\csname LT8\endcsname{\color{black}}%
    \fi
  \fi
  \setlength{\unitlength}{0.0500bp}%
  \begin{picture}(4680.00,3276.00)%
    \gplgaddtomacro\gplbacktext{%
      \csname LTb\endcsname%
      \put(1188,714){\makebox(0,0)[r]{\strut{}\tiny 10}}%
      \csname LTb\endcsname%
      \put(1188,1480){\makebox(0,0)[r]{\strut{}\tiny 100}}%
      \csname LTb\endcsname%
      \put(1188,2245){\makebox(0,0)[r]{\strut{}\tiny 1000}}%
      \csname LTb\endcsname%
      \put(1188,3011){\makebox(0,0)[r]{\strut{}\tiny 10000}}%
      \put(1254,352){\makebox(0,0){\strut{}\tiny 16}}%
      \put(1860,352){\makebox(0,0){\strut{}\tiny 32}}%
      \put(2466,352){\makebox(0,0){\strut{}\tiny 64}}%
      \put(3071,352){\makebox(0,0){\strut{}\tiny 128}}%
      \put(3677,352){\makebox(0,0){\strut{}\tiny 256}}%
      \put(4283,352){\makebox(0,0){\strut{}\tiny 512}}%
      \put(748,1747){\rotatebox{-270}{\makebox(0,0){\strut{}\tiny Time [s]}}}%
      \put(2768,154){\makebox(0,0){\strut{}{\tiny Number of processes} $\scriptstyle P$ {\tiny [-]}}}%
    }%
    \gplgaddtomacro\gplfronttext{%
      \csname LTb\endcsname%
      \put(3560,2893){\makebox(0,0)[r]{\strut{}\tiny local time}}%
      \csname LTb\endcsname%
      \put(3560,2783){\makebox(0,0)[r]{\strut{}\tiny global CG time}}%
      \csname LTb\endcsname%
      \put(3560,2673){\makebox(0,0)[r]{\strut{}\tiny total time}}%
    }%
    \gplbacktext
    \put(0,0){\includegraphics{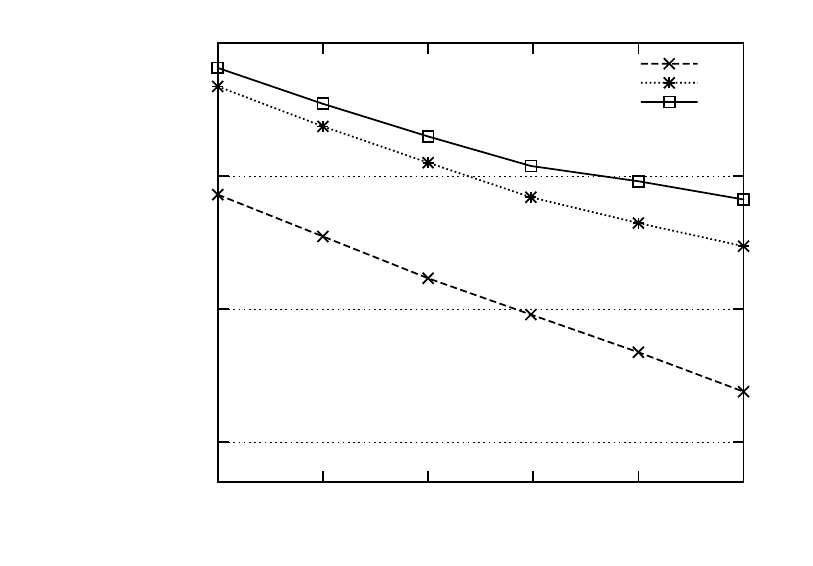}}%
    \gplfronttext
  \end{picture}%
\endgroup